\documentclass{article}

\usepackage{arxiv}

\usepackage[pdfborder={0 0 0}]{hyperref}
\usepackage[T1]{fontenc}
\usepackage{bm}
\usepackage{graphics,fancyhdr,graphicx,subfigure}
\usepackage{subfigure}
\usepackage{amsthm,amsmath,amsfonts,latexsym,amssymb}
\usepackage{lineno}
\usepackage{diagbox}
\usepackage[ruled,linesnumbered]{algorithm2e}
\SetKwComment{Comment}{$\triangleright$\ }{}
\usepackage{multirow,booktabs}
\usepackage{listings}
\usepackage[table]{xcolor}
\usepackage{lscape}
\usepackage{comment}
\usepackage{tabularx}
\usepackage[]{algorithm2e}

\usepackage{graphicx}
\graphicspath{{images/}}
\usepackage{tabularx}
\usepackage{nomencl}
\usepackage{array}
\usepackage{caption}
\usepackage{breqn}
\usepackage{lineno}
\usepackage{mathtools}
\usepackage{subfigure}
\usepackage{caption}
\usepackage{lmodern}
\usepackage{calligra}
\usepackage{xspace}
\usepackage{listings}
\usepackage{float}

\def\equationautorefname~#1\null{%
  Eq.~(#1)\null
  }
\def\subfigureautorefname~#1\null{%
  Fig.~#1\null
}

\definecolor{listinggray}{gray}{0.9}
\definecolor{lbcolor}{rgb}{0.9,0.9,0.9}
\definecolor{Darkgreen}{RGB}{0,100,0}

\title{Efficient hybrid topology optimization using GPU and homogenization based multigrid approach}

\author{
  Arya Prakash Padhi \\
  CTRANS\\
  Indian Institute of Technology Roorkee\\
  Roorkee - 110016, U.K., India \\
   \And
 Souvik Chakraborty \\
  Department of Applied Mechanics\\
  School of Artificial Intelligence\\
  Indian Institute of Technology Delhi\\
  Hauz Khas - 110016, New Delhi, India \\
  \texttt{souvik@am.iitd.ac.in} \\
   \And
 Anupam Chakrabarti \\
  Department of Civil Engineering\\
  Indian Institute of Technology Roorkee\\
  Roorkee - 110016, U.K., India \\
  \And
  Rajib Chowdhury \\
  Department of Civil Engineering\\
  Indian Institute of Technology Roorkee\\
  Roorkee - 110016, U.K., India \\
  \texttt{rajib.chowdhury@ce.iitr.ac.in}
}

\begin{document}
\maketitle

\begin{abstract}
We propose a new hybrid topology optimization algorithm based on multigrid approach that combines the parallelization strategy of CPU using OpenMP and heavily multithreading capabilities of modern Graphics Processing Units (GPU). In addition to that significant computational efficiency in memory requirement has been achieved using homogenization strategy. The algorithm has been integrated with versitile computing platform of MATLAB for ease of use and customization. The bottlenecking repetitive solution of the state equation has been solved using an optimized geometric multigrid approach along with CUDA parallelization enabling an order of magnitude faster in computational time than current state of the art implementations. On-the-fly computation of auxiliary matrices in the multigrid scheme and modification in interpolation schemes using homogenization strategy removes memory limitation of GPUs. Memory hierarchy of GPU has also been exploited for further optimized implementations.  All these enable solution of structures involving hundred millions of three dimensional brick elements to be accomplished in a standard desktop computer or a workstation.
Performance of the proposed algorithm is illustrated using several examples including  design dependent loads and multimaterial.Results obtained indicate the excellent performance and scalability of the proposed approach.
\end{abstract}


\section{Introduction}
\label{sec:intro}

Structural topology optimization attempts to find the efficient distribution of materials within a design domain under specific loading and boundary conditions. Unlike size and shape optimization, topology optimization allows for the creation of a material distribution without the need for a pre-determined structural arrangement. This gives engineers a strong tool for identifying creative and high-performance solutions ideas throughout the conceptual design phase and that's not even mentioning the huge impact that optimizing geometry and topology has on structural form.
Since Bendsoe and Kikuchi's \cite{bendsoe1988generating} early work on this problem, numerous research have been conducted in a wide range of physics issues, including stiffness maximization in structures, compliant mechanism design, and temperature maximization \cite{yoon2018multiphysics, zhu2019temperature, kazemi2020multi, chakraborty2019surrogate, giraldo2020multi}.

Density-based methods, implicit boundary moving methods (level set and phase field methods), and topological derivate based methods are the three major categories of shape and topology optimization methods \cite{sigmund2013topology}. In the first group, a fixed grid of finite elements is utilised to find the best void/solid material layout that minimises a specified objective function. The homogenization technique and Solid Isotropic Material with Penalization (SIMP) are two of the most common topology optimization methods existing in the literature. The second group of techniques includes those that employ implicit representations of structural boundaries. Such a boundary can be changed using the Level-Set Method (LSM) or phase field models by tracking the motion of a level-set function or altering the interfacial dynamics of phase field equations. The third group of methods is based on an explicit description of the structural form via a computational mesh or computer assisted design. In this work, density-based topology optimization techniques has been used because of its simplicity and wider use.

Topology optimization has made significant advances in theory and practise over the last decade, but computing needs continue to be a key impediment. Some complex activities, such as solving huge equation systems, estimating sensitivities, and choosing filtering strategies, are required in the topology optimization pipeline. As a result, if the model is big, topology optimization might take hours or even days \cite{sigmund2020eml}. Topology optimization necessitates high-performance computing (HPC), which addresses the challenge using task-level parallel computing.

It is becoming increasingly common to use graphics processing units (GPUs) for non-graphics applications, and this trend is only expected to increase in near future. As a result of its high computing capacity for Massive Parallel Processing (MPP) at a reasonable price, these graphics cards are being used in conjunction with a CPU to accelerate compute-intensive applications. Because of memory issues and lack of data-level parallelism, this is not an easy goal to achieve. Although the GPU's arithmetic processing engine is rapid, the memory from which this data is supplied may be slower. In addition to the non-coalesced global storage, shared memory access creating bank conflict, device (GPU) memory bandwidth is frequently insufficient. Simultaneous Single Instruction Multiple Data (SIMD) parallel computation, for which GPU architectures are designed, cannot be used due to the lack of data. Topology optimization methods must, therefore, be properly formulated and selected so that they can take full advantage of massively parallel architectures while avoiding memory problems, which severely limit GPU performance.
Still researchers have tried to overcome the challenges and GPU computing has been utilized successfully in a wide range of engineering and scientific issues that need numerical analysis.

The goal is to use data locality to decrease the computational cost of the topology optimization process. Wadbro and Berggren \cite{wadbro2009megapixel} advocated using GPU computation to evaluate high-resolution finite element models in heat conduction topology optimization in their early work. To decrease device memory needs, they used a Preconditioned Conjugate Gradient (PCG) technique with an assembly-free element-wise implementation. Schmidt and Schulz \cite{schmidt20112589} suggested a GPU-based nodal-wise assembly-free PCG solution for addressing elasticity issues at iterations of the minimization of the structural compliance problem with the SIMP technique.
The iterative solver's matrix-vector operations provided significant speed up by utilizing shared memory in the proposal. Reducing the grain size in the assembly-free GPU implementation is another way to boost GPU speed \cite{martinez2015fine}.

Multigrid techniques have benefited from the usage of GPU computing. These methods are among the most effective and widely used for resolving large linear equation systems. In structural mechanics Krylov subspace techniques frequently utilize these approaches as a preconditioner. The primary drawback of employing GPU computing to create geometric multigrid techniques is the amount of memory required to store the coefficient matrix and interpolation operators at various levels. Therefore, Dick et al \cite{dick2011real} developed an efficient nodal-wise matrix-free GPU implementation of the geometric multigrid technique with stencil computing for elasticity problems solved by the finite element method for this purpose. In order to avoid storing the coefficient matrix, they employ a Cartesian grid and parallel GPU computation to do ``on-the-fly'' calculations instead. By utilizing a stencil algorithm, data locality may be used as well.
This enables the merging of memory access into a single memory transaction and provides fast stencil-based ``on-the-fly'' grid transfer operators. Other notable work in this area includes \cite{schmidt20112589, martinez2015fine, dick2011real, baiges2019large, li2021topology, xie2020hierarchical}.

In spite of several research works efficient GPU implementation has still remained a challenging field due to asynchronous computational nature of CPU and GPU, memory size and bandwidth limitation of GPUs and non availability of software tools for easy integration and customization; 
the objective of this paper is to address some of these limitations.
To that end, we propose a hybrid scheme that exploits the optimal capacity of both GPU and CPU by appropriate distribution of the computations.
For example while most of the matrix-vector or matrix-matrix multiplications are performed in GPU, some of the additions that are required are done in OpenMP parallization in CPU. This takes advantages of the computational efficiency of the CPU and GPU architecture and make the end user program efficient. For further computational savings, we propose the use of multigrid algorithm; this results is significant computational efficiency as the solution to the fine level discretization is obtained by mapping the problem to a coarse grid resolution and solving it there and back-interpolating to the fine grid level. We further optimize it modifying these mapping and interpolation schemes based on a homogenization technique which requires even much less storage of the data. This facilitates the most parts of the computation to be carried out ~``on-th-fly''~ rather than retrieving the data from the computer memory and doing the operations. This essentially saves a significant portion of the remaining memory requirement of the variable storage. Finally the developed tool in integrated with TOP3D125 MATLAB code which gives easy understanding and seamless modifications several other aspects of the topology optimization algorithm like use of a different filtering scheme, different optimization routine, or even different physics equation altogether as long as it is in the form of $\mathbf K \bm u =\bm f $.       

The rest of the paper is organized as follows. In section 2 the standard density based topology optimization algorithm is described. In section 3 multigrid pre-conditioned conjugate gradient approach is explained. In the next section its efficient hybrid implementation is detailed. Subsequently several numerical experiments are carried out in section 5. Finally a homogenization based approach is described in section 6  and important observations are summarized in section 7.

\section{Density-based topology optimization}\label{sec:ps}

A binary programming problem, topology optimization aims to find the optimal material layout (solid and void) that minimizes an objective function \cite{bendsoe1989optimal}. When designing a material layout, it's important to adhere to a set of design constraints. As a result of their conceptual simplicity, density-based methods are the most widely used topology optimization methods in commercial or industrial software. Usually the design domain is discretized for two-fold benefit. First one is to update the density of each discretized element independent of others and the second one is to use the above discretization scheme to compute response of the system using finite element methods. It is possible to formulate the topology optimization problem as follows \cite{bendsoe2003topology}:

\begin{equation}
\label{eq:topopt}
\begin{split}
\min_{\rho} : \;\;\;\; & c(\bm \rho,\bm u) \\
\text{s.t} : \;\;\;\; & \mathbf K(\bm \rho) \bm u = \bm{f} \\
\;\;\;\; &  V(\bm \rho) \leq V_0\\
\;\;\;\; &  0 \leq \rho(x) \leq 1, {x} \in \mathbb R,
\end{split}
\end{equation}
where $c$ is the cost function, $\bm{\rho}$ denotes density design variables, $ \bm{u} $ represents the response of the system, $\mathbf K $ is the global stiffness matrix, $\bm{f} $ is the force vector, and $\bm{x}$ is array of discretized elements. The design domain is demarcated by $\mathbb R$ and the target volume of optimized shape $V(\bm{\rho})$ must be smaller than a prescribed value $V_0$. The unknown density parameters, $\rho(\bm{x})$, are utilized to adjust the finite element's stiffness in the regular mesh. Although discrete densities are desirable, use of continuous form helps in easy gradient computation and smooth transition at boundaries. In reality, this parameterization results in design with huge regions of intermediate densities that, while numerically ideal, are impractical to produce \cite{mlejnek1992some}.
Hence the density is modified to an artificial density form for computational convenience and drive the solution to binary 0 (void) or 1(solid material). Typically, this problem is handled utilizing implicit relaxation/penalization approaches, which force the topology design towards solid/void topology. The solid isotropic material with pennalization (SIMP) approach employs implicit penalization techniques through a power-law interpolation function  between void and solid to calculate the stiffness matrix of each element $\mathbf K_e$ similar to \cite{zhou1991coc} as follows:
\begin{equation}
\label{eq:K_mod}
\begin{aligned}
\mathbf K_e = \mathbf K_{min}+\bm \rho_e^p (\mathbf K_0-\mathbf K_{min}),
\end{aligned}
\end{equation}
where $\mathbf K_{0} $ corresponds stiffness matrix when an element is fully solid and $\mathbf K_{min} $ corresponds to minimum stiffness for least allowable density of an element. The later is provided to avoid singularity issue. Even though the use of material interpolation scheme allows for the creation of designs which are almost solid and void, they destroy the optimization problem's convexity thereby, increasing the risk of ending up in a local minimum. As a result, continuation methods are commonly used when solving optimization problems in order to avoid premature convergence to local minima. Continuation-based methods take ``global'' information into account and are more likely to ensure ``global'' convergence, or at the very least, convergence to better designs \cite{sigmund1998numerical}. 

To prevent numerical challenges and modelling issues such as mesh-dependency of solution and checker-board patterns, the topology optimization problem should also be regularized utilizing extra density field constraints. The sensitivity filter is employed in this study since it has been demonstrated in practice to be successful in providing mesh-independent solutions \cite{bourdin2001filters}. Furthermore, gradient filtering is motivated by continuum mechanics and may favour convergence of particular length scales over others, therefore hastening convergence. The sensitivity filter provides computational advantages because it is not included in the optimality criteria (OC) updating scheme loop.  

The sensitivity filter, as shown below, adds some kind of smoothing on the derivatives of the objective function as follows:
\begin{equation}
\label{eq:sensitivity filter}
\frac{\partial \hat c (\bm \rho)}{\partial \rho_e} = \frac{\sum_{i \in NB_e}w(x_i, x_e)\rho_i \frac{\partial c (\bm \rho)}{\partial \rho _i}}{\max {(\gamma, \rho_e)}\sum_{NB_e}{w(x_i, x_e)}}
\end{equation}
where $NB_e$ is an element's neighbourhood set, $w(x_i, x_e)$ is a weighting function, and $\gamma > 0 $ is a small number to prevent division by zero. In present work, the weighting function is defined as:
\begin{equation}
\label{eq:Neighbourhood}
\begin{aligned}
w(x_i,x_e) = \begin{dcases*}
        r-||x_i-x_e||  & if $||x_i-x_e||  \leq r$\\
        0 & if $||x_i-x_e||  \geq r$
        \end{dcases*}
\end{aligned},
\end{equation}
while an element's neighbourhood is defined as:
\begin{equation}
\label{eq:NB_dist}
\begin{aligned}
NB_e \coloneqq \{i \mid dist(i,e) \leq r\},
\end{aligned}
\end{equation}
where $r$ is the filter radius and $dist(i, e)$ emphasizes that it includes all elements $i$ within the distance $R$ from the center of element $e$.

Although SIMP can be used for solving a wide array of problems including heat sinks and other multi-physics problems \cite{zeng2019topology,sigmund1998topology}, we are interested in  minimization of structural compliance,

\begin{equation}
\label{eq:compliance objective}
\begin{aligned}
c = \bm{f}^{T}\bm{u},
\end{aligned}
\end{equation}
where $ \bm{f}$ is the applied force vector and $ \bm{u} $ is the corresponding displacement vector. 

Considering the discretized linear state system $\mathbf K \bm u = \bm f$ the sensitivities of Eq. \eqref{eq:compliance objective} using  adjoint state method (solving for $u^*$ in $\mathbf K \bm u^* = \frac{\partial c}{\partial \bm u}$)  with respect to $ \bm \rho $, we obtain
\begin{equation}
\label{eq:sensitivity}
\begin{aligned}
c_{\rho} = \frac{\partial c}{\partial \bm \rho} = -\bm u^{*T} \frac {\partial \mathbf {K}}{\partial \bm \rho} \bm{u} = -\bm u^{*T} (p\rho^{p-1}(\mathbf {K}_{0}-\mathbf {K_{min}})) \bm{u}.
\end{aligned}
\end{equation}

\noindent where $p$ is a penalty factor (usually 3 but can be obtained more accurately by continuation methods) and $(.)^T$ is the transpose of the vector. These sensitivities given by Eq. \eqref{eq:sensitivity} permit to update the design variables $\bm \rho$ using some sequential convex approximations, such as Sequential Quadratic Programming (SQP) \cite{wilson1963simplicial} or Method of Moving Asymptotes (MMA) \cite{svanberg1987method}. The Optimality Criterion (OC) updating scheme  proposed by \cite{bendsoe1995optimization} is adopted in this work due to its numerical efficiency. The OC updating scheme is as follows:

\begin{equation}
\label{eq:OCUp1}
\begin{aligned}
\rho_{e_{k+1}} = \begin{dcases*}
        \text{max} \{ ( 1-m) ,0 \}  & if $\rho_{e_k}B^{\eta}_{e_k}  \leq \text{max} \{ ( 1-m ),0 \}$,  \\
        \text{min} \{ ( 1+m ) ,1 \} & if \text{min} $\{ ( 1+m ),1 \}  \leq \rho_{e_k}B^{\eta}_{e_k}$, \\ 
        (\rho_{e_k}B^{\eta}_{e_k})^q & otherwise,
        \end{dcases*}
\end{aligned}
\end{equation}

\noindent where $ m $ is a positive move-limit, $\eta$ is a numerical damping coefficient (usually $\eta  = 1/2 $), q is a penalty factor to further achieve black-and-white typologies (typically $q = 2$ ) and

\begin{equation}
\label{eq:bisect}
\begin{aligned}
B_{e_k} = -\frac{\partial c(\rho)}{\partial \rho_e} \left( \lambda \frac{ \partial V(\rho)}{\partial \rho_e} \right )^{-1}
\end{aligned}
\end{equation}

\noindent is the Karush-Kuhn-Tucker (KKT) optimality condition. The Lagrange multiplier $ \lambda $ is found using bisection method. The algorithm stops when maximum number of iterations is reached or when the change in the variable $||\rho _{e_{k+1}} - \rho_{e_k} ||_{\infty}$ and change in objective function $ |c_{k+1} - c_k | $ fall below a  prescribed value.

The topology optimization pipeline's main bottleneck is the solution of the first constraint in Eq. \eqref{eq:topopt} which is typically calculated using finite element analysis (FEA). This stage entails two computationally demanding tasks: assembling the local element equations into a global system of equations and solving that resultant linear system. 
These computationally expensive tasks may result in an unsustainable situation in terms of computing time and memory usage. This issue is more prominent when working with large-scale models \cite{papadrakakis2011new} or when the system response must be re-evaluated again and again, as in topology optimization. Iterative solvers and assembly-free techniques have been widely utilized to reduce FEA memory requirement at the expense of increasing the processing time of the solution step, which is usually eased by parallel computing. This is explained in the next section.

\section{Multigrid pre-conditioned approach}\label{sec:sec3}

\subsection{Pre-conditioned Conjugate Gradient (PCG) Method}
The equation to be solved in the optimization loop discussed earlier is of the form of: 

\begin{equation}
\label{eq:lin_sys}
\begin{aligned}
\mathbf K \bm u =\bm f
\end{aligned}
\end{equation}

\noindent Efficient direct solution method involves cholesky or LU decomposition, where matrix $\mathbf K$ is decomposed into a lower triangular and an upper triangular matrix system and subsequently solved using forward and back substitution staying away from matrix inversion all throughout. However,  for large systems, this is still prohibitive and inefficient especially when matrix $\mathbf K$ is very large and a sparse one \cite{davis2006direct}.
The conjugate gradient technique is a mathematical methodology for numerically solving system of linear equations, specifically those with positive-definite matrices. It essentially solves Eq. \eqref{eq:lin_sys} but as a minimization problem of the following alternate quadratic form:

\begin{equation}
\label{eq:cg_q}
\begin{aligned}
f(\bm x) = \frac{1}{2}\bm u^{T}\mathbf K \bm u - \bm u^{T} \bm f.
\end{aligned}
\end{equation}

For positive definite system matrices (which is actually the case in many natural physical phenomena), the traditional Conjugate Gradient technique is the preferred iterative method \cite{hestenes1952methods}. It is used to minimize the functional $F(\bm{u})=\Vert \mathbf {K} \bm u- \bm f \Vert_{\mathbf {K}^{-1}}$ by multiplying the matrix vector just once in each iteration. As a matter of fact, this approach can theoretically arrive at the answer in less than $n$ iterations. In-fact the convergence rate can be given by,

\begin{equation}
\label{eq:condn_0}
\begin{aligned}
\Vert \bm u- \bm u_k \Vert_{\mathbf K^{-1}}  \leq  \Vert \bm u - \bm u_0 \Vert_{\mathbf K^{-1}} \left( \frac{\sqrt{\kappa} -1}{\sqrt{\kappa} + 1} \right)^k 
\end{aligned}
\end{equation}

\noindent where $\kappa$ is the matrix $\mathbf K$'s condition number, and $k$ is the number of iterations. It takes a long time for the system to reach convergence for $ \kappa > > 1$. Hence the original equation is usually modified for improved convergence by pre-multiplying both sides of Eq. \eqref{eq:lin_sys} by $\mathbf M^{-1}$. 

\begin{equation}
\label{eq:condn_1}
\begin{aligned}
\mathbf {M^{-1}K} \bm u = \mathbf M^{-1}\bm f,
\end{aligned}
\end{equation}
where $\mathbf M$ is a matrix or an operator such that $\kappa (\mathbf {M^{-1} K})<< \kappa (\mathbf K)$, and it is used to precondition the linear system in order to assure and accelerate convergence. 
There is a minimal expense involved in building an effective preconditioner $(\mathbf M^{-1})$ and the condition number should be as near to unity as feasible and independent of the number $n$. Incomplete Cholesky factorization, diagonal scaling, and Factorized Sparse Approximate Inverses (FSAI) are examples of classical preconditioners that do not offer mesh-independent convergence rates. In this study, contemporary multilevel/multigrid approach which is numerically scalable is used instead (e.g., \cite{vassilevski2008multilevel}). The PCG algorithm is depicted in algorithm \ref{alg:PCG}.

\begin{algorithm}[ht!]
\caption{ Preconditioned Conjugate Gradient Method}
\label{alg:PCG}
\textbf{Initialization}: $ x_0 = 0, r_0 = b, z_0 = M^{-1}r_0 = 0, k=1,  p_0 = z_0, \alpha_0 = r_0^Tr_0/(p_0^TAp_0) $\\
\While {$r_k^{T}r_k > $ tol} {
    $k = k+1$ \\
    $x_{k+1} = x_k + \alpha_kp_k$ \\
    $r_{k+1} = r_k - \alpha_kAp_k$ \\
    $z_{k+1} = M^{-1}r_{k+1} $\\
    $ \beta_{k+1} = r_{k+1}^Tz_{k+1}/(r_kz_k) $\\
    $p_{k+1} = z_{k+1} + \beta_{k+1}p_k$\\
    $\alpha_{k+1} = r_{k+1}^T z_{k+1}/(p_{k+1}Ap_{k+1})$ 
    
    }
\textbf{return $\bm x_{k+1}$}
\end{algorithm}

Multigrid Preconditioned Conjugate Gradient (MGCG) method has been successfully used in the discretized finite element solution part to solve the state space equation. This high-efficiency iterative technique has been used for solving large-scale linear equation because of its highly efficient preconditioning technique, faster convergence and minimal computational effort involment and scalability. The multigrid method upon which the preconditioner is built is described below.

\subsection{Multigrid Method (MG)}
Solution of the linear system of equation $\mathbf K \bm u = \bm f$ by classical iterative scheme, is generally done by resolving $\mathbf K$ into matrices $\mathbf M$ and $\mathbf N$ with non-singular $\mathbf M$, such that $\mathbf K = \mathbf {M-N}$. 
Thus,
\begin{equation}
\label{eq:MG_M_N}
\begin{split}
    \mathbf M \bm{u} & =  \mathbf N \bm{u} +\bm  f \\
    & \text{or} \\
    \bm{u} & = \underbrace{\mathbf M^{-1}\mathbf N}_{S} \bm{u} + \mathbf M^{-1} \bm f
\end{split}
\end{equation}
Given an initial iterate $\bm u^{(0)}$, a fixed point iteration can be applied to this equation

\begin{equation}
        \bm{u}^{(m+1)} = \mathbf S\bm{u}^{(m)} + \mathbf M^{-1} \bm f,\;\; m=0,1,2,3, \ldots
\end{equation}

This basic iterative approach might also be damped with damping coefficient $\omega$:

\begin{equation}
        \bm{u}^{*} = \mathbf S \bm{u}^{(m)} + \mathbf M^{-1} f, \,\,\,  \bm{u}^{(m+1)} = \omega \bm{u}^{*} + (1-\omega ) \bm{u}^{(m)}
\end{equation}
such that
\begin{equation}
    \bm u^{(m+1)}=(\omega \mathbf S+(1-\omega)\mathbb I) \bm u^{(m)}+\omega \mathbf M^{-1} \bm{f}
\end{equation}
If $\bm{u}$ is the actual solution of the original equation and $\bm{u}^{(m)}$ is the approximation computed using above, the error is denoted by,

\begin{equation}
        \bm{e}^{(m)} = \bm{u} - \bm{u}^{(m)}
\end{equation}
and the residual is given by

\begin{equation}
\label{eq:residual}
        \bm{r}^{(m)} = \bm{f} - \mathbf A \bm{u}^{(m)}
        \\ \Longrightarrow \mathbf A \bm{e}^{(m)} = \bm{r}^{(m)}
\end{equation}
When $ \mathbf M = \text{diag} (A) = D $ in Eq. \eqref{eq:MG_M_N}, a straightforward calculation can show that, the iterative solution is of the form:

\begin{equation}\label{eq:jacobi}
    \begin{aligned}
        \bm u^{(m+1)} = \bm u^{(m)} + D^{-1}\bm r^{(m)}
    \end{aligned}
\end{equation}
This is known as the Jacobi method. With damping coefficient $\omega$, the damped Jacobi method can be written as:
\begin{equation}
    \begin{aligned}
        \bm u^{(m+1)} = \bm u^{(m)} + \omega D^{-1}\bm r^{(m)}, \,\,\, \omega \in (0,1].
    \end{aligned}
\end{equation}
In multigrid (MG) methods, the residual calculation step (Eq.
\eqref{eq:residual}) is used for updating the current iterate $\bm u^{(m)}$. An approximation $\tilde{\bm e}^{(m)}$ of $\bm e^{(m)}$
is computed from Eq. \eqref{eq:jacobi} and the new iterate is given by $\bm u^{(m+1)} =\bm  u^{(m)} + \tilde{\bm e}^{(m)}$. 

Another important component of MG method is the grid transfer mechanism. A simple way for the same is depicted below in Fig \ref{fig:restr} in which a uniform refinement step consists of dividing in halves all intervals of domain $\Omega^{2h}$ in order to obtain the domain of  $\Omega^h$. 
Consequently, we solve $\mathbf K^h \bm u^h=\bm f^h,$ with $\bm u \in \Omega^h$. Projection of error from $\Omega^h$ to $\Omega^{2h}$ is called restriction (denoted by $ I_{h}^{2h})$ and that from $\Omega^{2h}$ to $\Omega^h$ is called prolongation or interpolation (denoted by $ I_{2h}^h)$

\begin{figure}[htb!]
        \centering
        {
        \includegraphics[width = 0.80\textwidth]{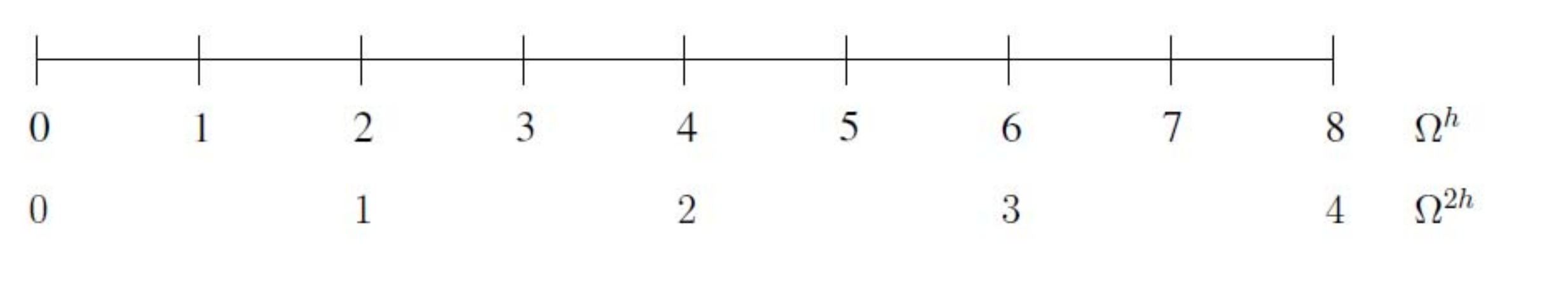}}\hspace{2mm}
        \caption{Coarse and fine grid}
        \label{fig:coarse-fine-grid}
        
\end{figure}
\begin{figure}[htb!]
        \centering
        {
        \includegraphics[width = 0.80\textwidth]{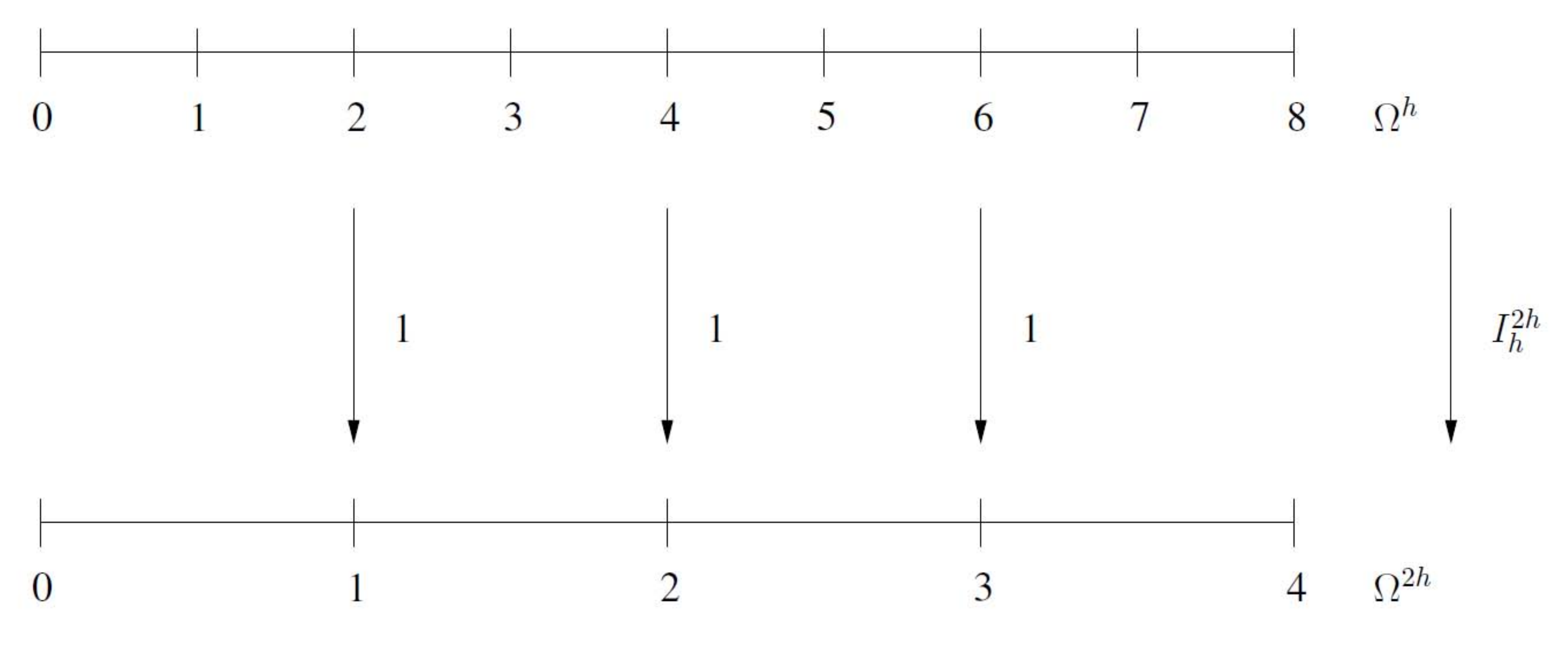}}\hspace{2mm}
        \caption{A simple restriction Operation }
        \label{fig:restr}
        
\end{figure}

\begin{figure}[h]
        \centering
        {
        \includegraphics[width = 0.80\textwidth]{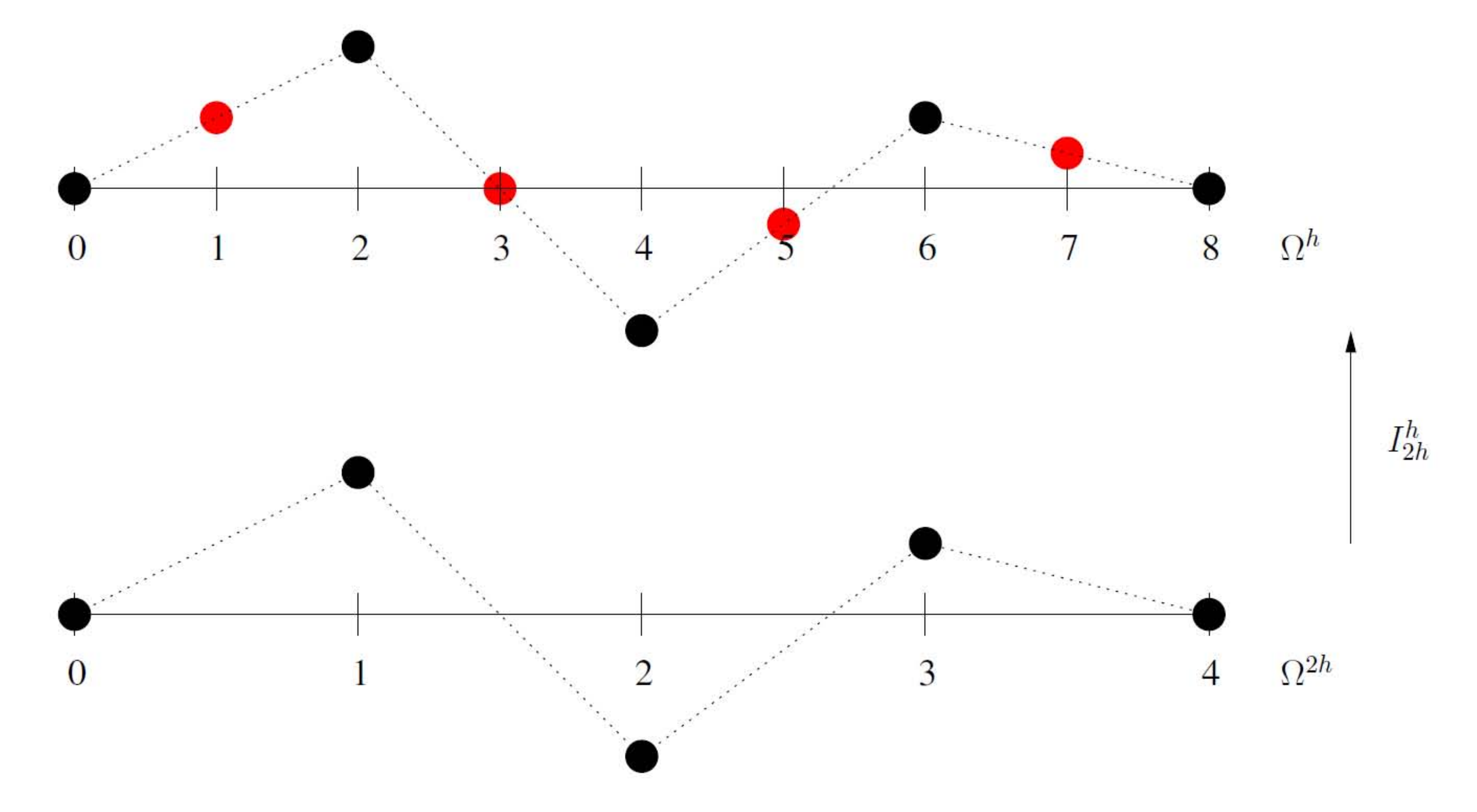}}\hspace{2mm}
        \caption{Linear interpolation operation}
        \label{fig:linint}
        
\end{figure}

The iterative single level methods (e.g Jacobi solution) can quickly reduce the high frequency components of the mistake, but it performs poorly for low frequency errors. 
The MG technique is well known for being one of the most efficient strategies to enhance the convergence rate. The goal of MG is to build multiple grids at different scale (resolution) of discretization. Then, at each level, repeated relaxations are performed to reduce high-frequency errors on tiny grids and low-frequency errors on coarse grids.
A linear system can be solved with the help of MG at a cost of $O(n)$. Smoothing and coarse-grid correction are two complimentary procedures that work together to produce optimal performance. Gauss-Seidel or Jacobi stationary iterative methods are commonly used to smoothing out the solution and decreasing oscillation errors. The combination of smoothing, restriction, and prolongation works very well and results in converged solution. A simple two grid algorithm is depicted in algorithm \ref{alg:2grid}. This algorithm basically transfers the equation from fine grid of $\Omega^h$ domain to the two times coarser grid of $\Omega^{2h}$, solves it there and interpolates back to the finer grid. 

\begin{figure}[h]
        \centering
        {
        \includegraphics[width = 0.50\textwidth]{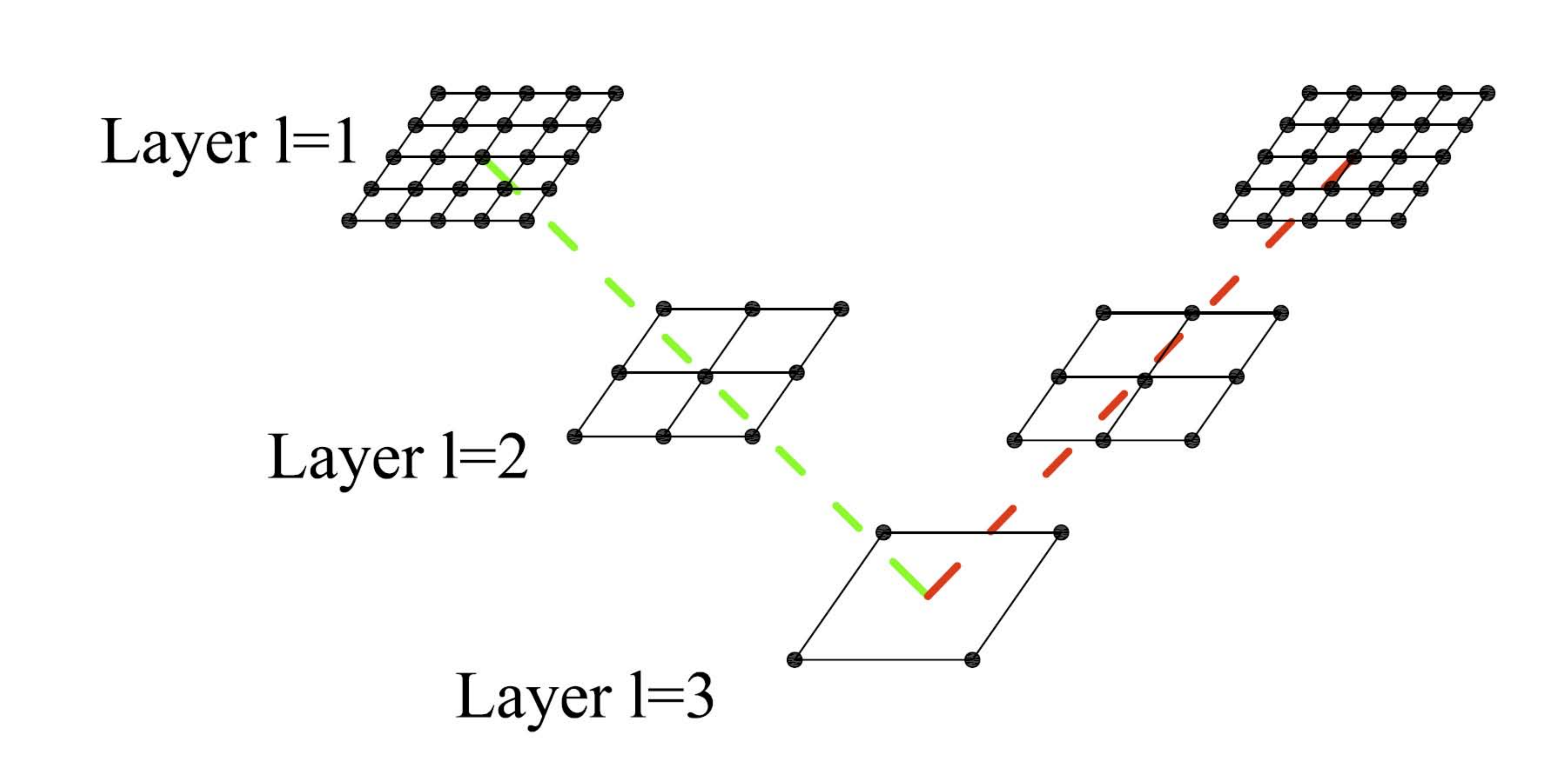}}\hspace{2mm}
        \caption{Multigrid levels from coarse  to fine}
        \label{fig:multiscale}
        
\end{figure}

\begin{figure}[htb!]
        \centering
        {
        \includegraphics[width = 0.50\textwidth]{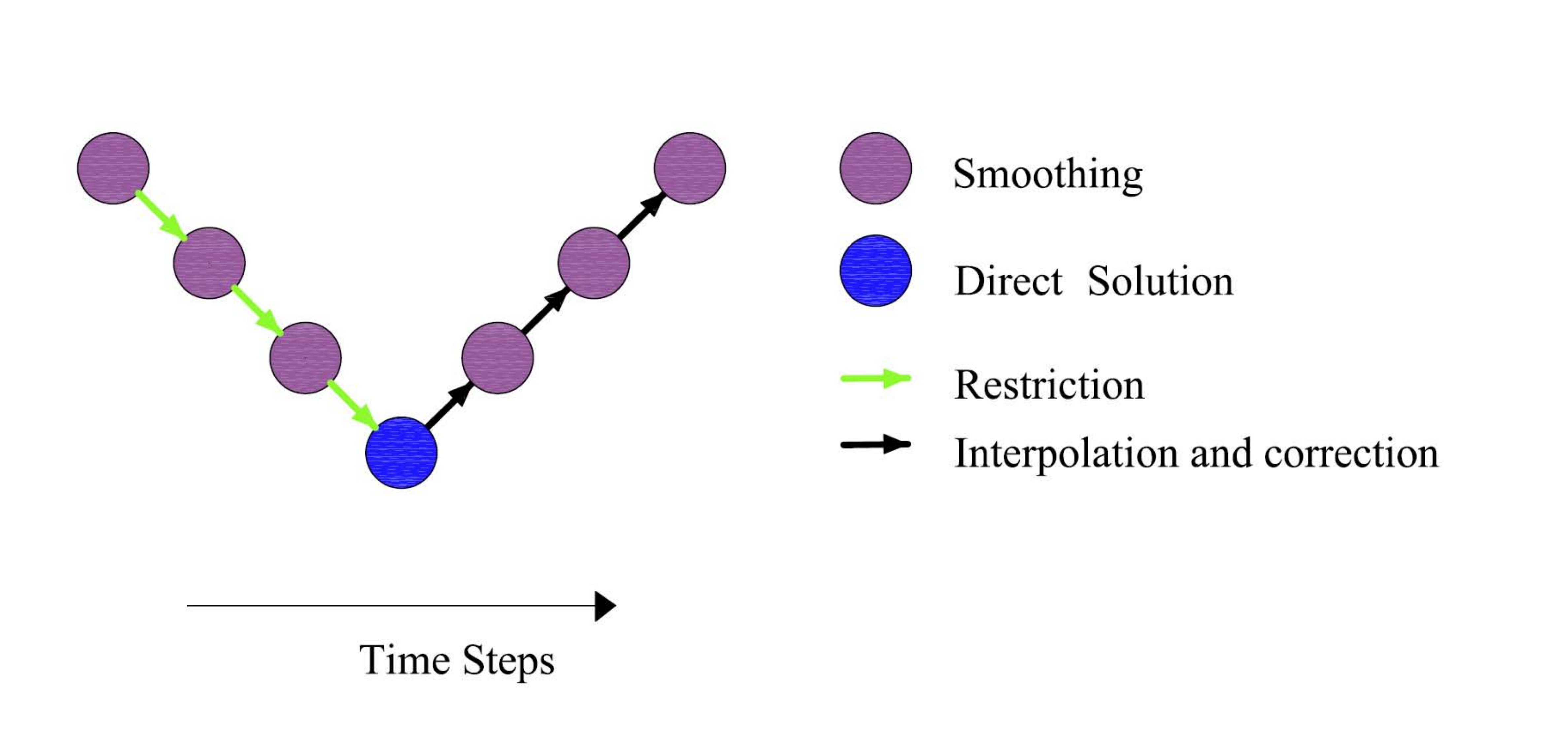}}\hspace{2mm}
        \caption{Schematic diagram of V-cycle with various operations}
        \label{fig:V-Cycle}
        
\end{figure}

The important question now, is the solution in coarser grid which is step 4 of algorithm 2. Looking carefully it is again essentially a linear system and can be solved in a manner we started solving in algorithm 2. This necessitates the repeated application of two-grid algorithm and gives rise to recursive hierarchical MultiGrid solution. This typical flow of processes is shown in Fig. \ref{fig:V-Cycle}, which is commonly known as V-cycle. The corresponding algorithm is written in algorithm \ref{alg:vgrid}. This approach splits the grids into multiple sizes and computes the precise answer only at the coarsest grid corresponding to the largest discretization size. In this multilayer mesh method, the finest mesh level is denoted by $l$ (level) = 1, whereas $l$ (level) = $L$ represents a coarser level as we can see in Fig. \ref{fig:multiscale}. This popular V-cycle method is utilized in this study. The combination of multigrid with conjugate gradient method simplifies the computational complexity of the problem significantly and its comparison with direct solver with number of degrees of freedom N is shown in Table \ref{tab:exp1_B2D}. Storage requirement as well as computational time is proportional to $N$ in mgcg but in direct solver it is $N\sqrt{N}$.

\begin{figure}[H]
        \centering
        \subfigure[]
        {
        \includegraphics[width = 0.40\textwidth]{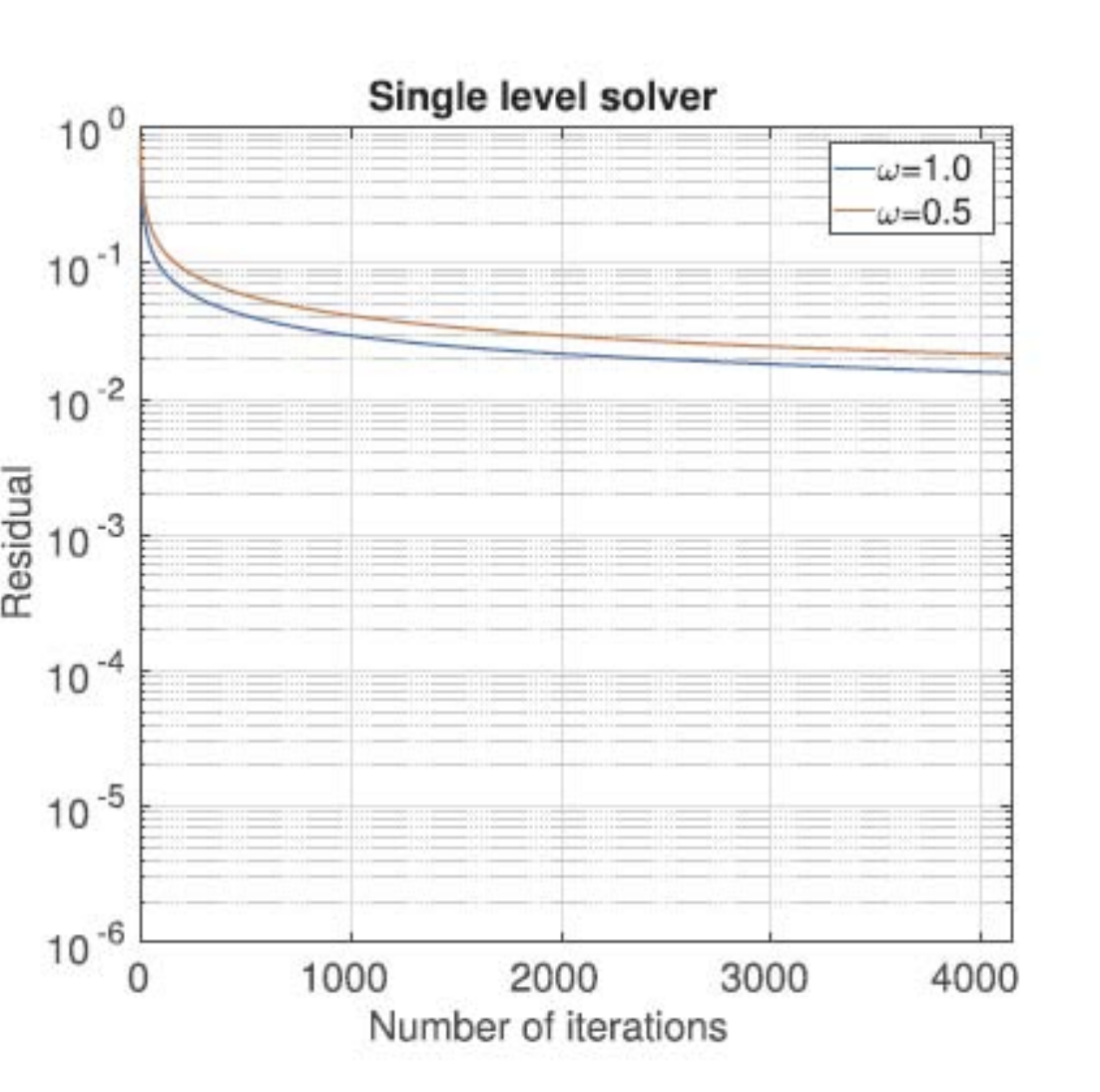}}
        \hspace{2mm}
        \subfigure[]{
        \includegraphics[width = 0.40\textwidth]{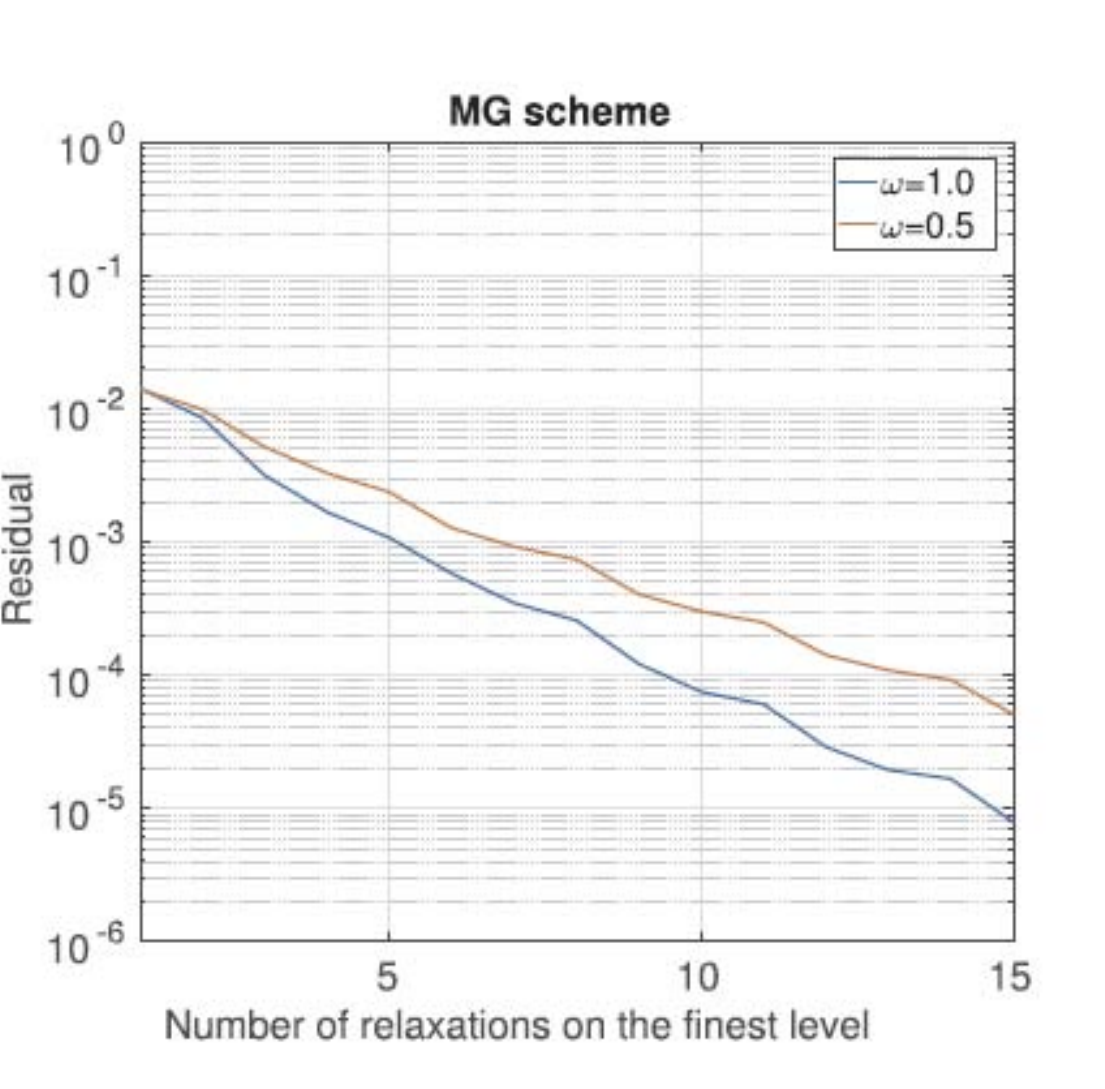}
        }
        \caption{(a) Single level (b) Multigrid Convergence}
        \label{fig:Single_Grid Vs MultiGrid}
        
\end{figure}

\begin{table}[!hbt]
    \centering
    \caption{MGCG vs Direct Solver}
    \label{tab:exp1_B2D}
    \begin{tabular}{|p{4cm}|p{3cm}|p{3cm}|}
        \hline 
            \textbf{Criteria} & \textbf{MGCG} & \textbf{Direct Solver} \\
        \hline
            Memory &  $N$ & $N\sqrt{N}$ \\
        \hline
        
            Computational Time &  $N$ &  $N\sqrt{N}$\\
        \hline
        
            Precision &  Approximate & Precise\\
        \hline
    \end{tabular}
\end{table}

\begin{algorithm}[H]
\caption{ Two-grid algorithm \textbf {u} =  MG\textbf{(u, f, K, h, S)}}
\label{alg:2grid}

\textbf{Pre-smooth:} $\bm u^h = \bm {u}^h+\bm S^{-1} \bm (f^h - \mathbf K^h \bm u^h) \,\,\,\, \text{on} \,\,\, \Omega^h$ \\
\textbf{Residual:} Compute Residual $\bm r^h = \bm f^h - \mathbf K^h \bm u^h$ \\
\textbf {Restriction: } Projection of $\bm r^h$ to $\bm r^{2h} \text{ to } \Omega^{2h}$. $ \bm r^{2h} = I_h^{2h} \bm r^h$ \\
\textbf{Solution on coarse Grid:} solve on $\Omega^{2h}. \,\,\, \mathbf K ^{2h} \bm e^{2h} = \bm r^{2h}$\\
\textbf{Interpolation:} Projection of $e^{2h} \text{ to } \Omega^h. \,\,\, e^h = I_{2h}^h e^{2h} $ \\
\textbf{Update:}  $\bm u^h = \bm u^h + \bm e^h$ \\
\textbf{Post-smooth:} $\bm u^h = \bm {u}^h+\bm S^{-1} \bm (f^h - \mathbf K^h \bm u^h) \,\,\,\, \text{on} \,\,\, \Omega^h$ \\
\textbf{return:}  $\bm u $ \\

\end{algorithm}

\begin{algorithm}[ht!]
\caption{ Multigrid Method with V-cycle $u^h \leftarrow MG_v(u^h,f^h, S, h ) $ of equation $\mathbf K^h \bm u^h = \bm f^h$}
\label{alg:vgrid}
\textbf{Pre-smoothing}: $\bm u^h = \bm {u}^h+\bm S^{-1} \bm (f^h - \mathbf K^h \bm u^h) \,\,\,\, \text{on} \,\,\, \Omega^h$ \\
\uIf{$\Omega^h $ is the coarsest grid}{Solve the problem directly} 
\Else{Restrict to next coarser grid:  $ \bm r^{2h} = \mathbf I_h^{2h} ( \bm f^h - \mathbf K^h \bm u^h)$ \\
Set initial iterate on next coarser grid: $\bm u^{2h} = 0 $\\
Call the V-cycle scheme one time for next coarser grid: $\bm u^{2h} \leftarrow MG_v(\bm u^{2h} ,\bm f^{2h})$}

\textbf{Prolongation correction}: $\bm u^h = \bm u^h + \mathbf I_{2h}^h \bm u^{2h} $ \\
\textbf{Post smoothing}: $\bm u^h = \bm {u}^h+\bm S^{-1} \bm (f^h - \mathbf K^h \bm u^h) \,\,\,\, \text{on} \,\,\, \Omega^h$ \\
\textbf{Return}: $\bm u $\\
\end{algorithm}

\section{GPU implementation of SIMP method}\label{sec:GPU_impl}
MG techniques, although relatively efficient, also suffers from the curse-of-dimensionality; this is particularly true when dealing with real-life systems having millions of degree-of-freedom. In this section, we propose a hybrid implementation of the SIMP method that exploits both GPU and MPI programming. We first discuss the CUDA architecture for GPU implementation followed by the proposed framework.
\subsection {CUDA Architecture}
\begin{figure}[htbp!]
    \centering
    {
    \subfigure[]{\includegraphics[width = 0.60\textwidth]{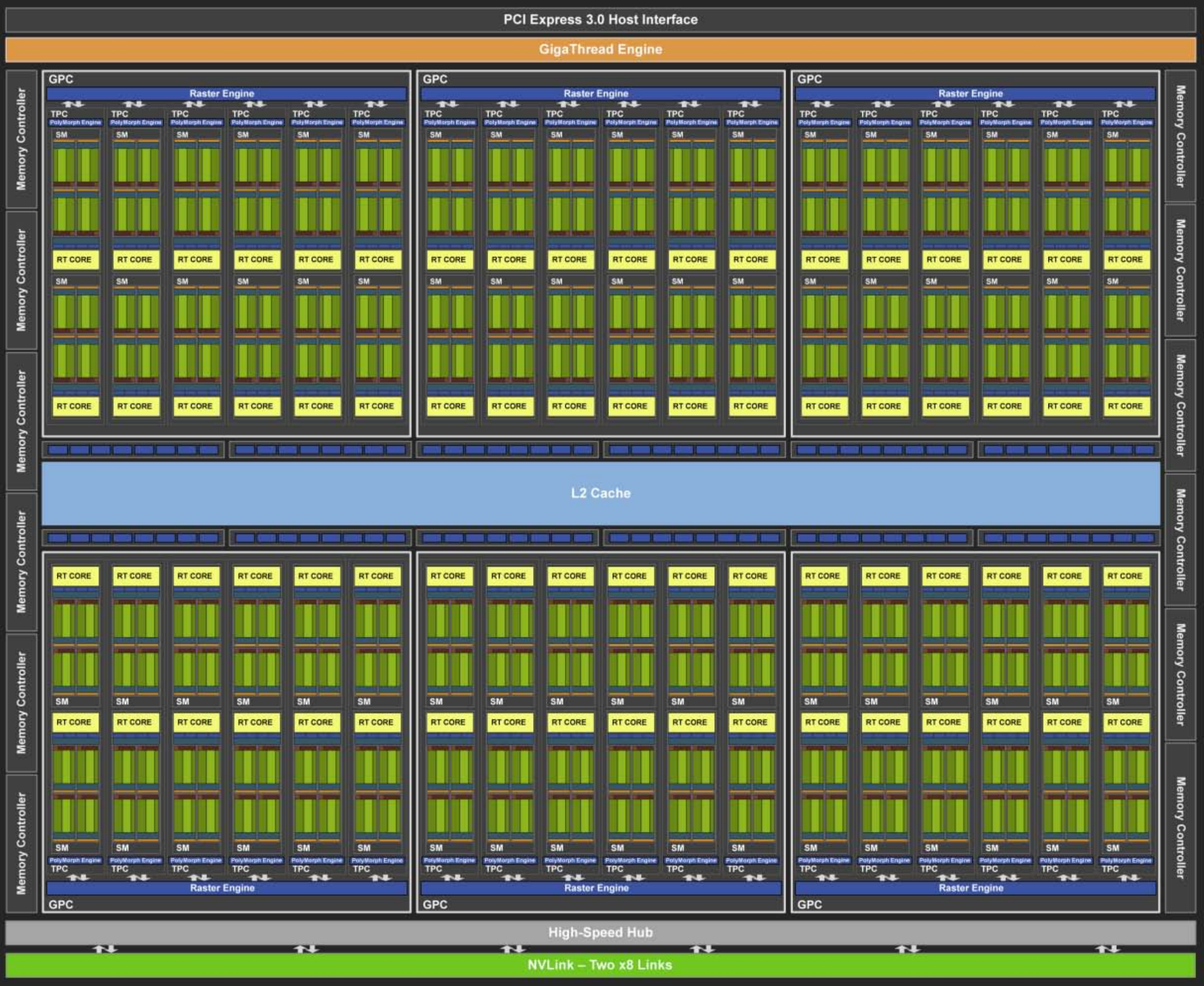}}
    \subfigure[]{\includegraphics[width = 0.29\textwidth]{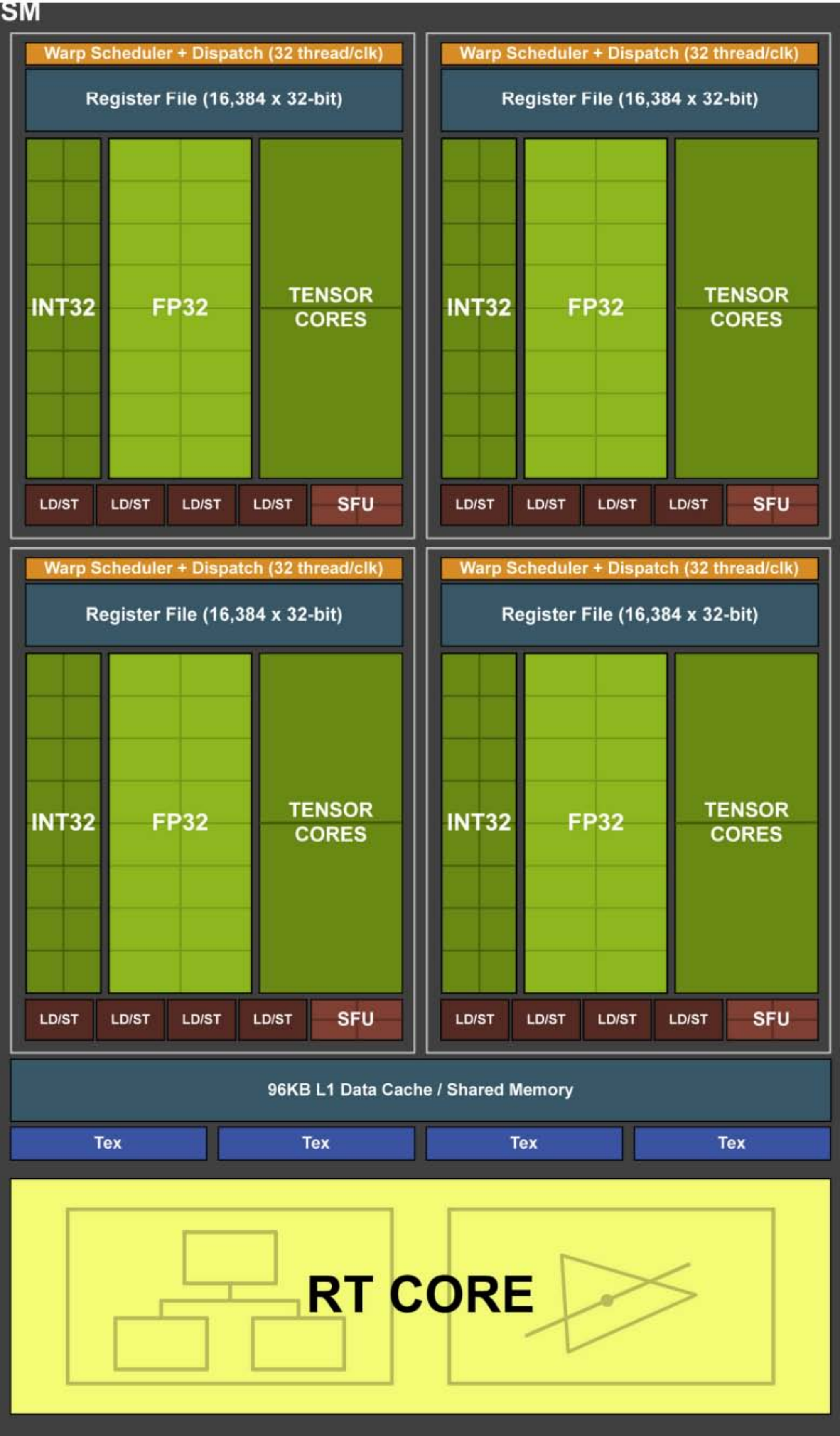}}
    \subfigure[]{
    \includegraphics[width = 0.80\textwidth]{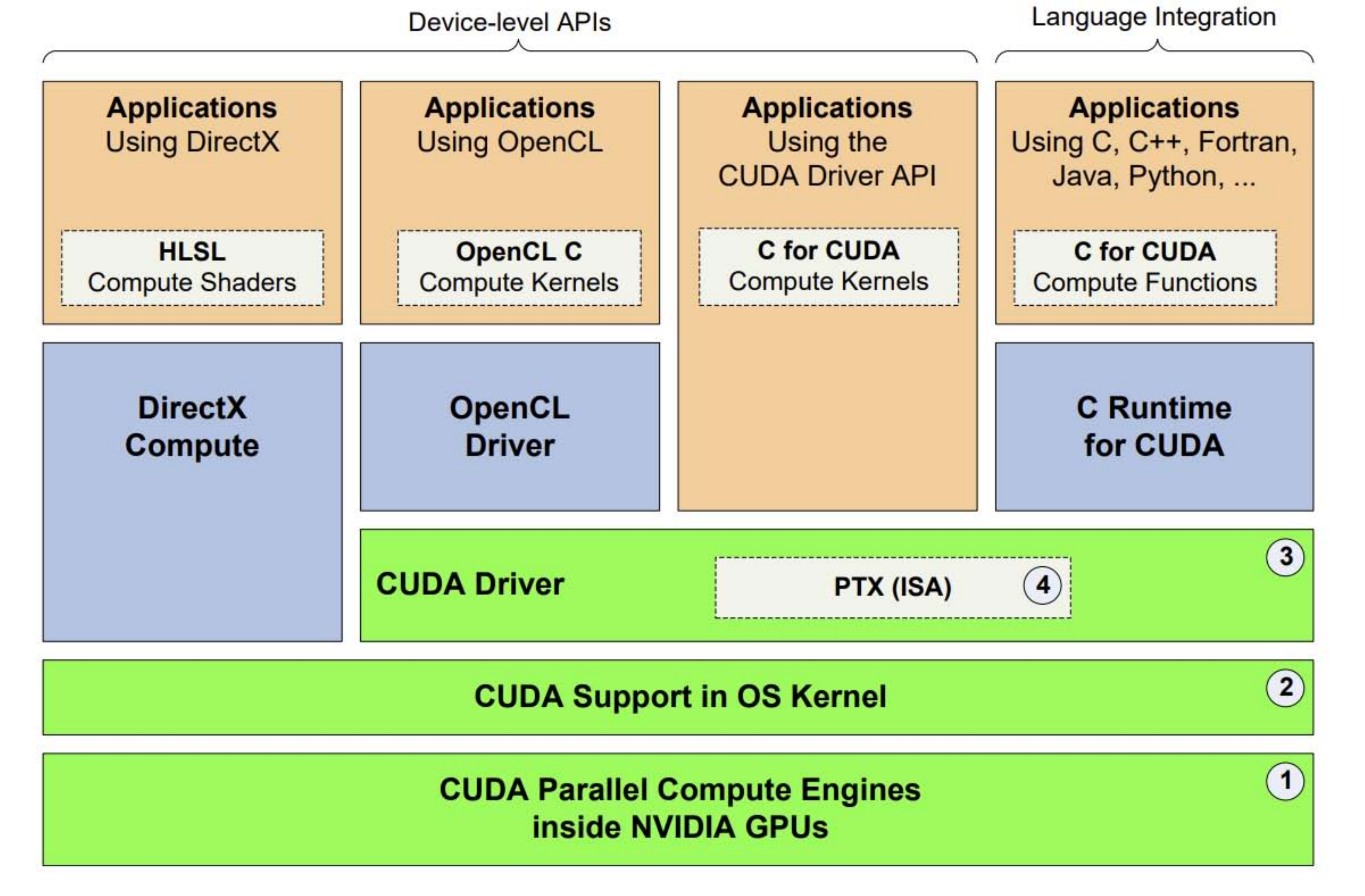}}\hspace{2mm}
    \caption{CUDA GPU Architecture}
    \label{fig:GPU Architecture}
    }
\end{figure}

GPUs were created to meet the market's need of fast and realistic 3D rendering in real time. Their tremendous computational capability at a reasonable cost is making them increasingly attractive in non-graphics HPC applications. The schematic diagram of a modern turing architecture GPU in depicted in figure \ref{fig:GPU Architecture}(a) and \ref{fig:GPU Architecture}(b) below. Compute Unified Device Architecture (CUDA), a programming paradigm developed by Nvidia, is currently the most widely used GPU programming model. So-called data-parallel computing (data/SIMD parallelism) can be performed on the GPU by leveraging several processor cores. ``Kernel'', a C Language Extension function, is used to run the parallel code (one instruction, multiple data). According to, the kernel call should indicate the number of CUDA threads structured as a grid of thread blocks, as seen in figure \ref{fig:GPU Architecture}(a) \cite{NvidagpuGduide113}. The software level hierarchy is depicted in Fig. \ref{fig:GPU Architecture}(c) \cite{NvidagpuGduide113}.

The CUDA Architecture consists of numerous components (Fig. \ref{fig:GPU Architecture}), including (i) NVIDIA GPU's Parallel computation Engines (computation blocks), (ii) OS hardware initialization support, (iii) Kernel-level support, (iv) User mode driver providing developers with a device level API, (v) Set of parallel computing functions and functions via PTX instruction architecture (ISA). We exploit first four feature in the implementation strategy described below.

\subsection{Hybrid Topology Optimization Framework}
\begin{algorithm}[t!]
\caption{ Topology Optimization (Eq. \eqref{eq:topopt}), \\Given data: design domain geometry and discretization (\text{ nelx, nely, nelz}), loading ($\bm f$ )and boundary condition ($\text{fixeddofs}$),  target volume fraction}
\label{alg:simp}
\textbf{Initialize}: Initialization of empirical parameters $ \bm \rho, p, r, m, \eta, \text{ch}=1,  k=0 $ \\
\textbf{Stiffness Matrix}: Compute $\mathbf K_0$ \\
\textbf{Prepare filter}: Compute $\bm w$ \Comment*[r]{Eq. \ref{eq:Neighbourhood}, \ref{eq:NB_dist}}
\While{\text{ch} > 0.01} 
{
    Compute modified stiffness $\mathbf K_e $ \Comment*[r]{Eq. \ref{eq:K_mod}}
    Direct solve for $\bm u \text{ in } \mathbf K_e \bm u = \bm f $ \\
    Sensitivity computation $\frac{\partial c}{\partial \bm \rho}$ \Comment*[r]{Eq. \ref{eq:sensitivity}}  
    Filtering of sensitivities   \Comment*[r]{Eq. \ref{eq:sensitivity filter}} 
    Update $\bm \rho_{k+1} $ from $\bm \rho_{k} $  \Comment*[r]{Eq. \ref{eq:OCUp1}, \ref{eq:bisect}} 
    $k=k+1$ \Comment*[r]{loop count} 
}
\textbf{Return}: The final $\bm \rho$ and visualization\\
\end{algorithm}

We revisit the standard topology optimization algorithm using SIMP method (Algorithm \ref{alg:simp}). We note that step 6 corresponds to majority of the computational cost. One obvious solution is to leverage the powerful CUDA platform. However, CUDA has limited memory and hence, shifting the computation to CUDA will compromise the scalability of the algorithm. To address this issue, we propose a computing framework that leverage the strength of both GPU and Open Multi-Processing using CPU cores. 

\begin{figure}[!h]
    \centering
    {
    \includegraphics[width = 1.0\textwidth]{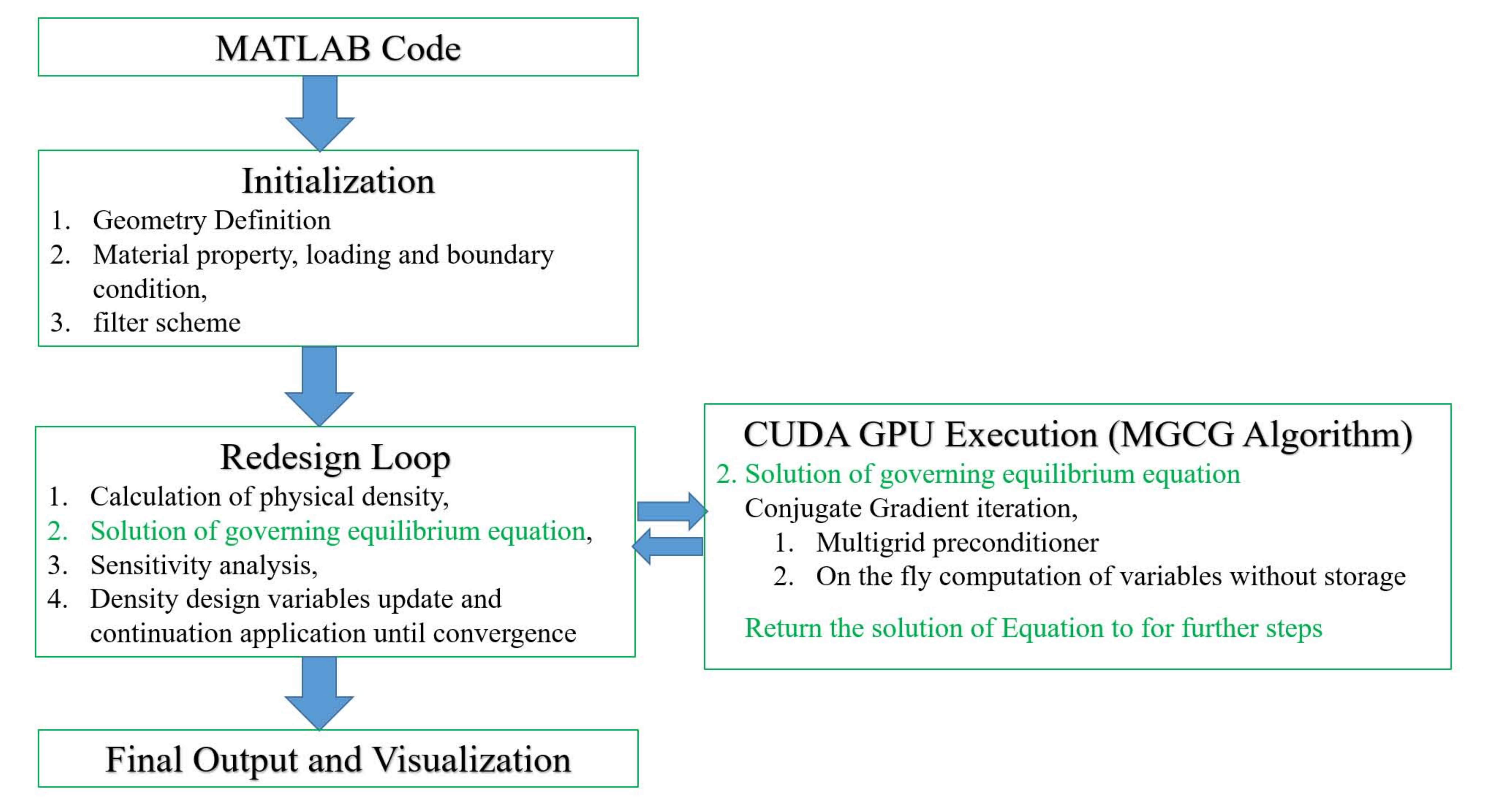}}\hspace{2mm} 
    \caption{Topology Optimization Framework}
    \label{fig:topopt_flowchart}
    
\end{figure}

The proposed topology optimization framework has multiple components. First and foremost, we replace the direct solver in step 6 of Algorithm \ref{alg:simp} with Multigrid preconditioned conjugate gradient (MGCG) \cite{amir2014multigrid}. 
To further optimize the process, the MGCG solver's computation has been offloaded to GPU's highly efficient computational architecture. A 3-Dimensional MATLAB code for MGCG-based minimal compliance topology optimization has been developed here using the density based SIMP formulation. The computational intensive linear equation solver part in redesign loop of topology optimization is written in CUDA C language and compiled using NVIDIA nvcc compiler and called from MATLAB backbone using it's mex function capability. It is worthwhile to mention that the program developed utilizes both GPU and CPU cores. To be specific, the computationally expensive operations like matrix multiplication and matrix inversions are carried out using GPU. For clarity of the readers, the broad framework fo the proposed approach is shown in Fig. \ref{fig:topopt_flowchart}. The beauty of the developed framework is that the mex function call can be used as a black box to solve any linear system of equations of the form $\mathbf K \bm u =\bm  f $ in cartesian discretization with 8-noded brick elements. Subsequently, it can be trivially incorporated within other topology optimization algorithms as well.

\begin{figure}[!h]
    \centering
    {
    \includegraphics[width = 1.0\textwidth]{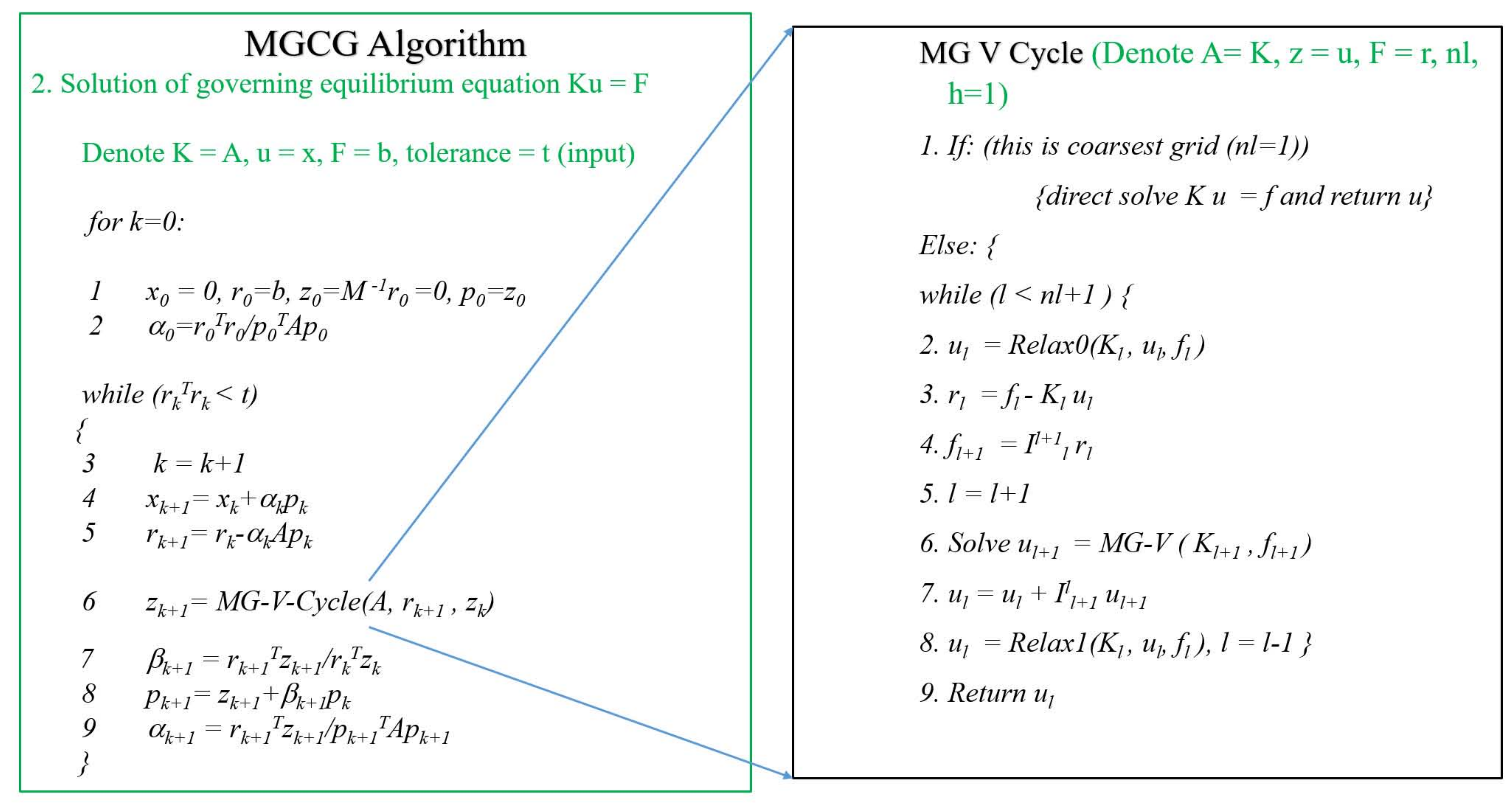}}\hspace{2mm}
    \caption{CUDA (MGCG) Framework}
    \label{fig:MGCG-V-Cycle}
\end{figure}

Next, we shift our focus to the MGCG algorithm within the proposed algorithm. Although MGCG is significantly efficient as compared to a direct solver, it still accounts for majority of the computational cost. To accelerate this step, we divide it into two parts as shown in figure \ref{fig:MGCG-V-Cycle}; the first one corresponds to the conjugate gradient loop and the second one corresponds to the multigrid preconditioning inside the conjugate gradient loop.
Preconditioning enhances the convergence rat and multigrid based preconditioning improves overall stability of the algorithm. The MATLAB code calls the compiled MGCG CUDA code using the mex function capability. Also, the main loop of MGCG algorithm is controlled on CPU memory. But the important variables are synchronized to GPU memory as well because the matrix multiplications and additions in steps 2 to 9 of MGCG algorithm in Fig. \ref{fig:MGCG-V-Cycle} are to be carried out in GPU in an heavily multithreaded environment. The addition and subtraction operation of step 3 and 4 can be carried out in CPU only as these don't require huge multithreading. But flow of data from GPU device memory to host CPU memory adds time to overall process making it little less efficient. Hence these are also carried out in GPU. Similarly the main loop of multigrid V-cycle (which is basically step 6 of MGCG main loop) is controlled in CPU which then subsequently calls GPU kernel functions for the operations in steps 1 to 7 etc.

Although GPU multithreading is highly recommended, sometimes unavailability of NVIDIA GPU or incompatibility issue prevent from execution of task. Hence two separate functions are developed, one is TopOptMGCGOMP which is a purely CPU multi-threaded version (with no GPU related hardware and software requirement), and it utilizes open-source Open Multi-Processing(OpenMP) package to implement efficient CPU multi-threading. Another one is the hybrid TopOptCUDA version, which uses NVIDIA GPU using CUDA development package and CPU. Efficiency of both versions depend upon the relative strength of the CPU and GPU pair and numerical comparison between both the parts are given in section 5. The challenge in the CUDA version is that although GPU's have tremendous multithreading potential it's native memory storage is quite limited and if it has to access data from host CPU memory, a significant time is spent on data transfer from CPU to GPU readable memory. Hence effective strategy has been developed to store as little data as possible in GPU memory without hindering the computation itself. This is described in Section 4.2.1. Also a simple homogenization technique is shown to facilitate further significant reduction in storage requirement in Section 4.2.2.  

\subsubsection{Use of local stiffness Matrix for various computation}
Instead of using assembled stiffness matrix for each and every computation, majority of the computation has been done with local stiffness matrices. This essentially saves precious GPU memory used for extremely fast computation. Each node inside of design domain is common to 8 elements, so sum of density times local stiffness matrices of these 8 elements gives the value equal to global stiffness value of that node. This strategy helps to perform nodal computations of all matrix and vector product and addition without explicit storage of global assembled matrices. In previous two subsections while calculating residue for each iteration in MGCG similar strategy can be seen to be in use. Although this increases number of floating point operation by roughly two times this technique decreases memory consumption significantly.

\subsubsection {Homogenization Strategy}
To optimize the memory requirement, we utilize the well-known homogenization strategy. This is a simple yet effective technique to further optimize memory requirements. Traditionally while moving from a fine grid for the equation to coarse grid in V-cycle, restriction operation is used on $\bm f$ and $\mathbf K$ for transformation $\bm f^{2h} = \mathbf I_h^{2h} \bm f^h$ as shown in Fig. \ref{fig:linint}. The restriction operation is generally a Galerkin approximation given by equation:

\begin{equation}
\begin{aligned}
\mathbf I_h^{2h} =  \frac{1}{4}
\begin{bmatrix}
1 & 2 & 1 &    &    &    \\
  & 1 & 2 & 1  &    &    \\
  &   & 1 & 2  & 1  &    \\
  &   &   & ... &   &    \\
  &   &   &     & ... &  \\
  &   &   & 1   & 2   &  1
\end{bmatrix},
\end{aligned}
\end{equation}
and,
\begin{equation}
\begin{aligned}
    \mathbf K^{2h} =  \mathbf I_h^{2h} \mathbf K^h \mathbf I_{2h}^h.
\end{aligned}
\end{equation}
This type of restriction leads to storage of local element stiffness matrices at coarser level. One one hand this eliminates the need for on-the-fly computation; however, the memory requirement increases. To reduce the memory requirement, we employ a homogenization scheme where we density of the coarse element is computed as mean of density values of eight neighboring elements at the finer level; this enables on-the-fly computation of the local stiffness matrix and reduces the need for storage.
\subsubsection {Memory}
Suppose a structural design domain has $n$ degrees of freedom. Referring to MGCG algorithm in Fig. \ref{fig:MGCG-V-Cycle}, the memory cost of the GPU can be divided into two parts; one corresponding to storage requirement in the preconditioned conjugate gradient (PCG) loop and the other part is in the V-cycle. There are four unique variables ($P, Q(=AP), R \text{ and } Z)$ in the PCG iterations and five others in the V-cycle iterations. Each of these variables are of dimension $n \times 1$. So the total storage comes out to be $4n + 5n = 9n$. But in the V-cycle loop, the variables are also required to be stored in in each successive coarser levels each having $1/8$ times the size of previous finer level. Thus, the storage requirement due to the V-cycle part can be increased by $20 \%$ and hence, the total storage requirement is $4n + 1.2 \times 5n = 10n$. For clarity of readers, the above calculation is summarized in Table \ref{tab:GPU_variables}.

\begin{table}[!hbt]
    \centering
    \caption{Storage requirement in GPU memory}
    
    \label{tab:GPU_variables}
    \begin{tabular}{|p{2cm}|p{3cm}|p{3cm}|p{2cm}|}
        \hline 
            \textbf{Loop} & \textbf{Vectors} & \textbf{Dimension of each vector} & \textbf{Sub Total}  \\
        \hline
            PCG &  $ P, Q, R, Z $ & $n\times 1$ & 4n\\
        \hline
        
            V-cycle &  $U, F, R, CX, AD $  &  $n\times 1$ & 5n\\
        \hline \hline
            \textbf{Total} & \multicolumn{3}{r|}{$4n + 1.2 \times 5n = 10n$} \\

        \hline
    \end{tabular}
\end{table}

To further illustrate the memory requirement, we consider a structural system having 100 millions degrees of freedom. We further assume that each variable has double precision. With this setup, the storage requirement will be approximately equal to $100 \times 10^ 6 \times 10 \times 8 \,\,\, bytes  = 8 \text{GB}$.
Decreasing the precision will even increase the capability of solving bigger size problems. Even with double precision it is quite modest considering 100 million elements can be fit into a standard desktop GPU having 8GB of memory these days. 
\section{Numerical Experiments}\label{sec:Nu_Ex}
In this section, we present four numerical examples to illustrate the efficacy and robustness of the proposed approach. The examples are arranged in the increasing order of complexity and involves real-life scenarios such as bridges, buildings, and design dependent loading. We compare the results obtained with those obtained using the state-of-the-art topology optimization algorithms currently available (TOP3D125 and GPU based topology optimization code \cite{wu2015system}). Computational time, no. of iterations, and average memory required have been considered as comparison metrics. Finally, to illustrate the versatility of the proposed approach, the developed topology optimization framework is tested on three computational environment: (a) standard GPU workstation with 8GB VRAM  (System 1), (b)  old GPU workstation with 2GB VRAM (System 2), and (c) regular GPU laptop with 4GB VRAM  (System 3). The detailed specifications of the computational environmental is described in the appendix.
\subsection{Example 1: 3D cantilever beam}
\begin{figure}[htbp!]
        \centering
        \subfigure[]{
        \includegraphics[width = 0.25\textwidth]{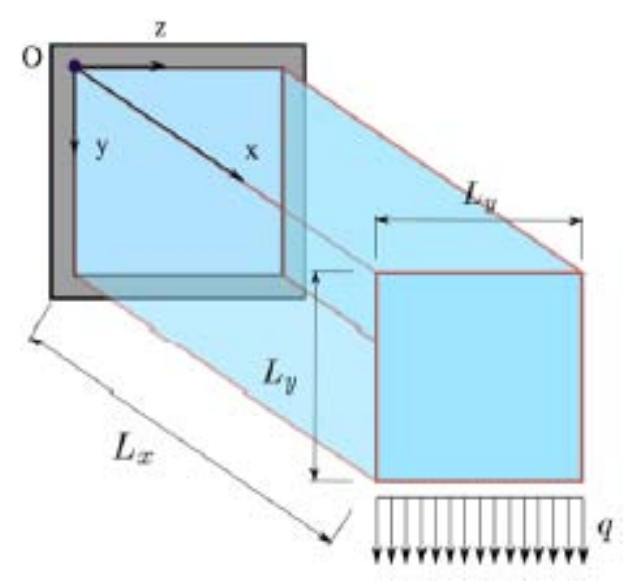}}
        \hspace{8mm}
        \subfigure[]{
        \includegraphics[width = 0.25\textwidth]{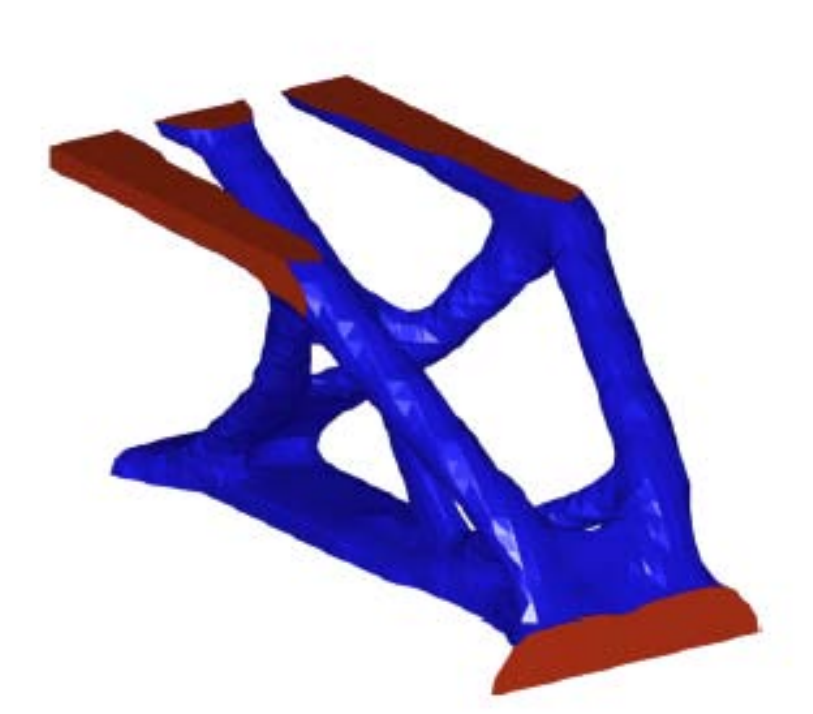}}
        \subfigure[]{
        \includegraphics[width = 0.3\textwidth]{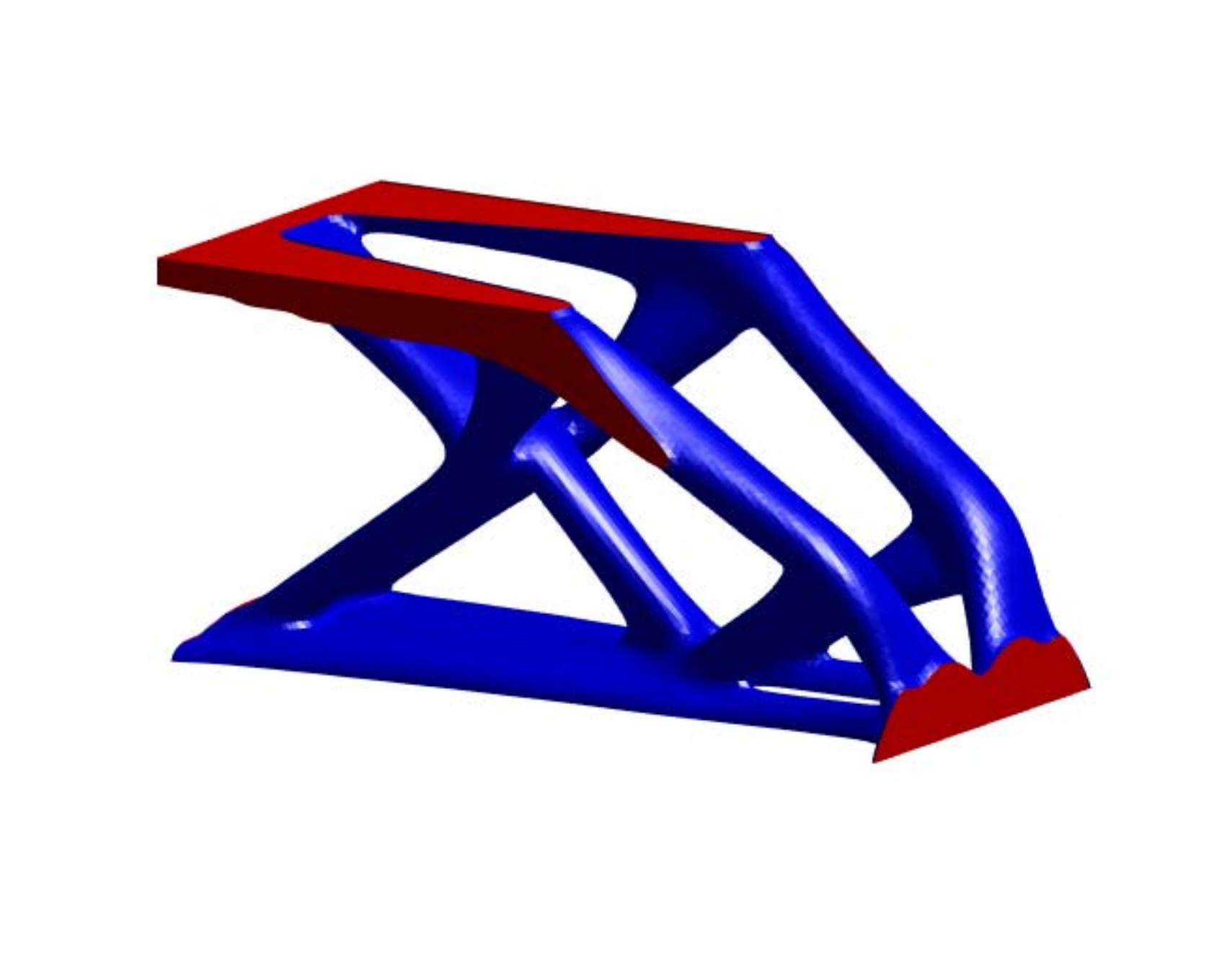}}
        \caption{(a) Cantilever Beam Problem Description (b) Result obtained from \cite{ferrari2020new} (c) Result obtained from our code for 64x32x32 discretization}
        \label{fig:cant_loading}
\end{figure}

As the first example, we consider a 3D cantilever beam subjected to line load $q$. The loading, boundary condition and design domain are shown in Fig. \ref{fig:cant_loading}. As already stated, the objective here is to minimize the structural compliance. We consider two separate cases, one where the design domain is discretized into $64 \times 32 \times 32$ elements and another where the design domain is discretized into $128 \times 64 \times 64$ elements. Subsequently, the two cases have $65,536$ and $524,288$ design variables.

\begin{figure}[htbp!]
        \centering
        \subfigure[]{
        \includegraphics[width = 0.45\textwidth]{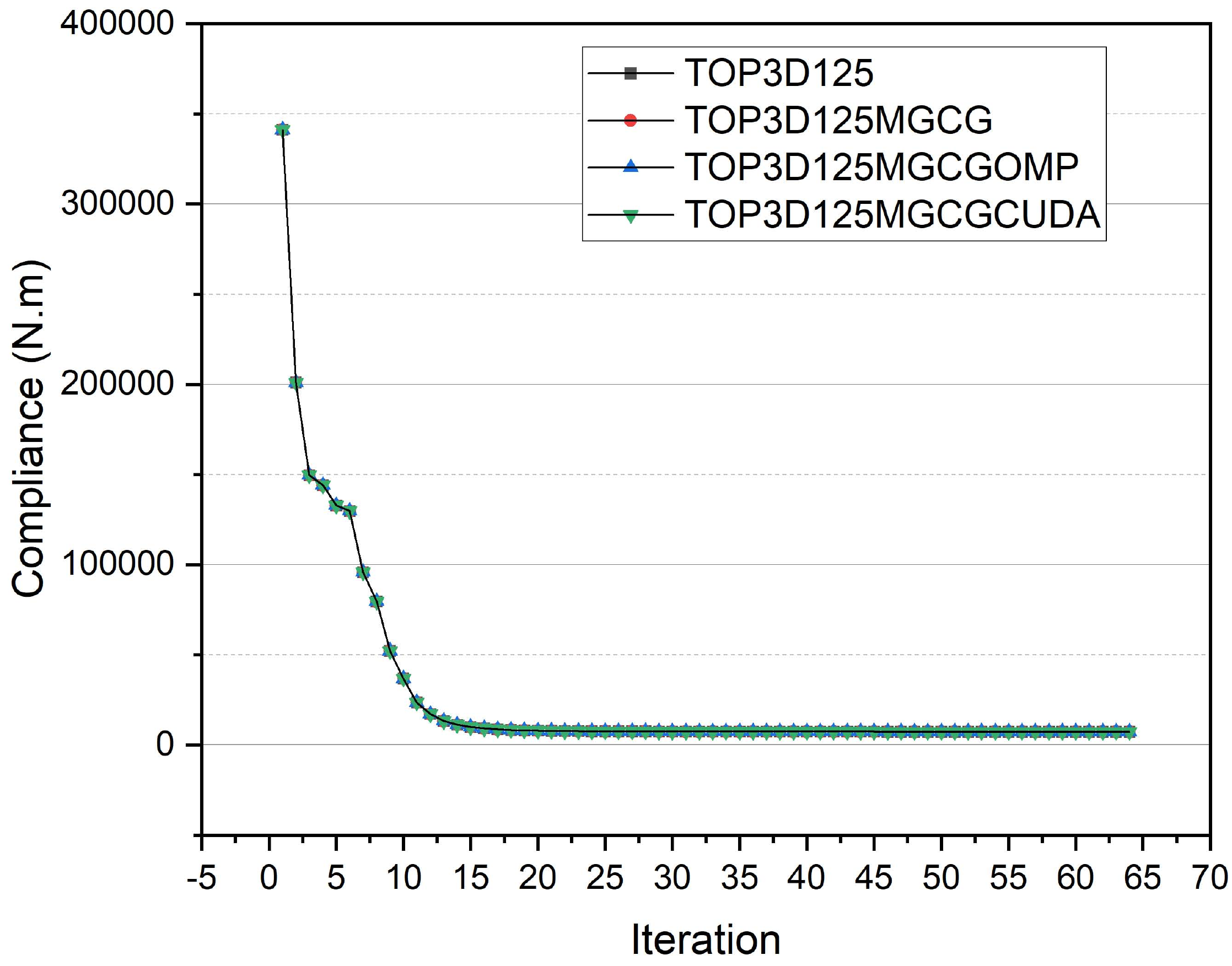}}
        \hspace{2mm}
        \subfigure[]{
        \includegraphics[width = 0.45\textwidth]{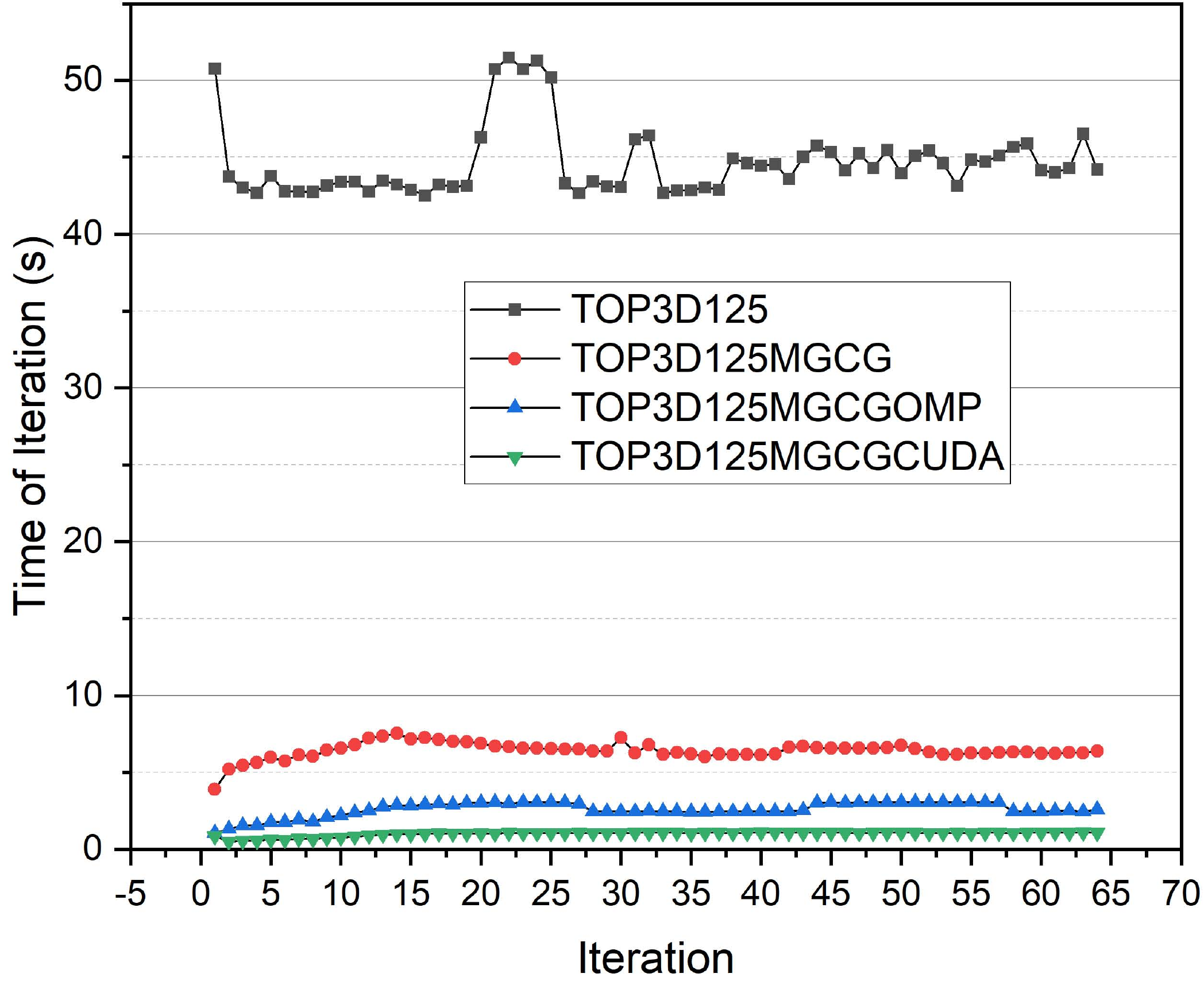}}
        \subfigure[]{
        \includegraphics[width = 0.45\textwidth]{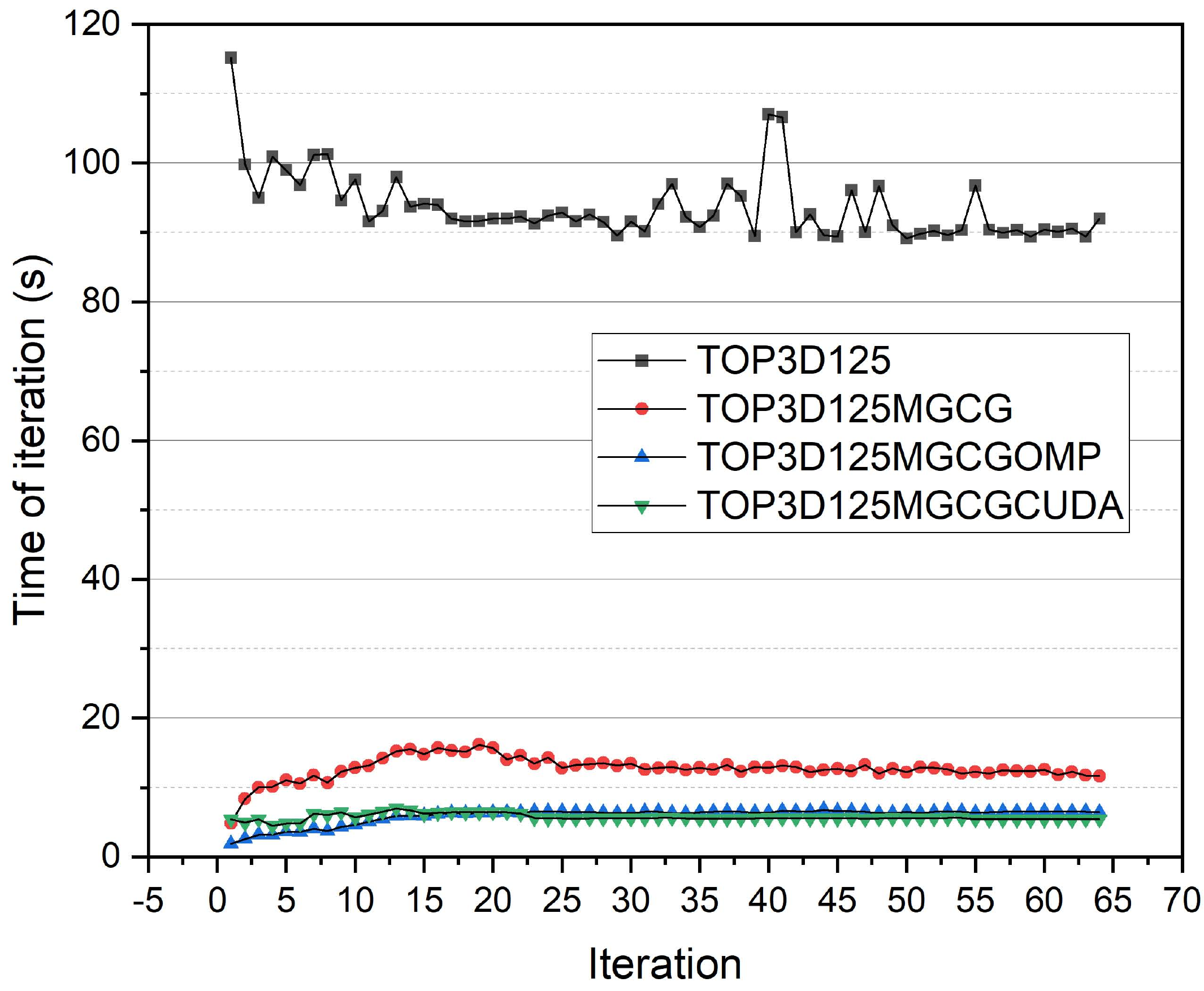}}
        \subfigure[]{
        \includegraphics[width = 0.45\textwidth]{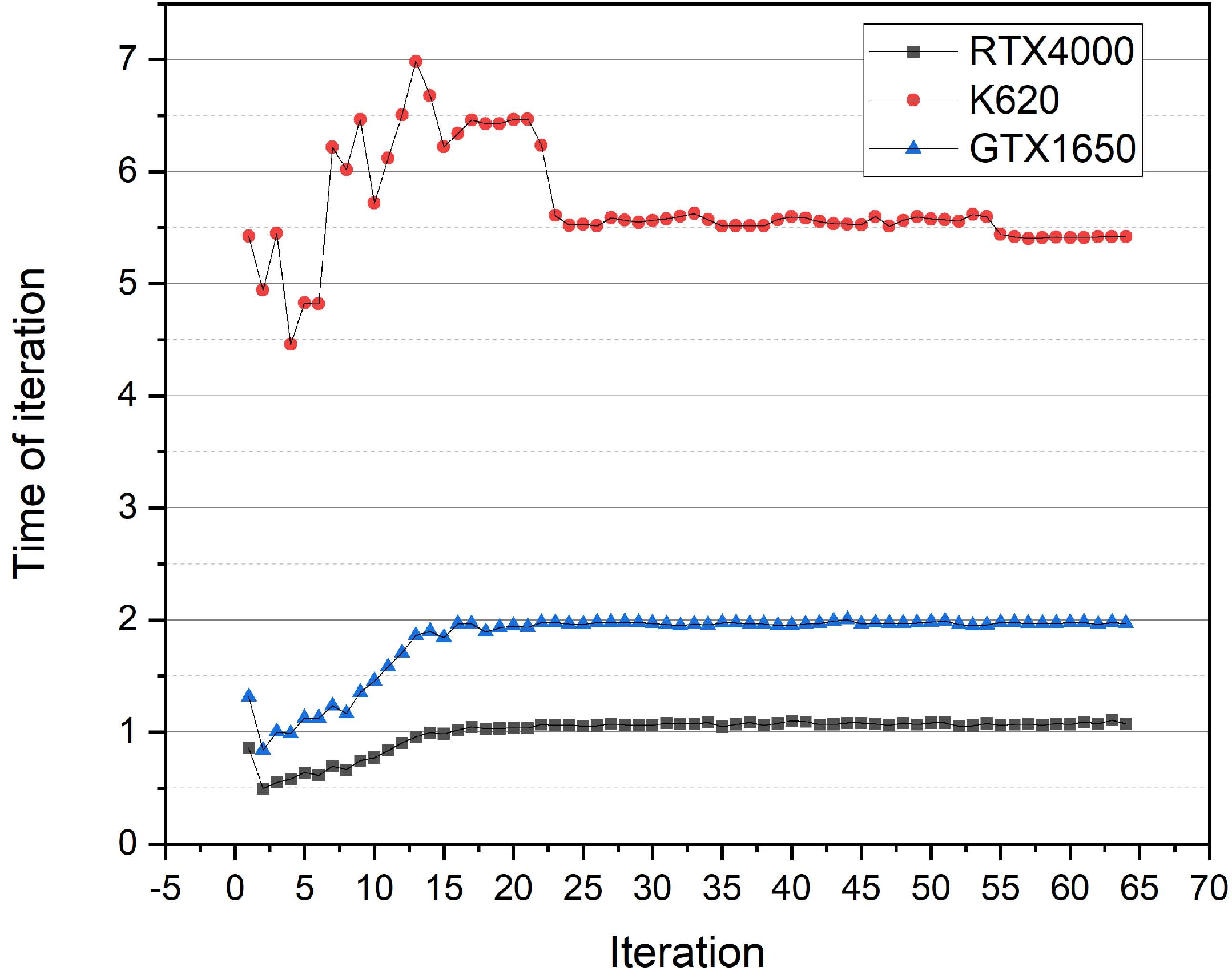}}
        \caption{(a) Cantilever Beam (64 x 32 x 32)  Convergence of objective function (b) Time of iterations in system 1 (c) Time of iterations in system 2 (d) Comparison of CUDA codes in 3 systems}
        \label{fig:cant_conv}
\end{figure}

Fig. \ref{fig:cant_loading} (b) and (c) show the optimal topology obtained using the method proposed in \cite{ferrari2020new} and the proposed approach for grid size of $64 \times 32 \times 32$. We observe that the results obtained using the proposed approach matches exactly with that obtained using \cite{ferrari2020new}; this essentially validates the accuracy of the proposed approach. For illustrating the efficiency of the proposed approach, we compare the computationally efficiency and memory requirements. 
Table \ref{tab:com_sys1} depicts the comparison of the proposed approach with the state-of-the-art topology optimization framework proposed in \cite{ferrari2020new}. Additionally we show the computational gain achieved by including only multigrid preconditioned iterative solver (TOP3D125MGCG), multigrid preconditioned iterative solver parallelized using OMP (TOP3D125MGCGOMP), and multigrid preconditioned iterative solver parallelized using GPU and CUDA (TOP3D125MGCGCUDA). We observe that TOP3D125MGCGCUDA on the standard workstation is approximately 45 times faster as compared to the optimized code presented in \cite{ferrari2020new}. As compared the OMP version of the code, the CUDA version proposed is about 3 times faster. The gain in computational memory is even more significant with the proposed approach. While the TOP3D125 \cite{ferrari2020new} requires 15.08 GB memory, the proposed TOP3DMGCGOMP and TOP3DMGCGCUDA require only 3.27, 2.60, and 2.03 GB memory only. It is noteworthy to mention that the memory for TOP3DMGCGCUDA represents GPU memory. The advantage of the proposed framework becomes more prominent for grid size of $128 \times 64 \times 64$. The state-of-art Top125 fails in this case due to huge memory requirement. The MGCG based frameworks, one the other hand, yields satisfactory results with minimal increase in the memory requirements. The CUDA based framework is approximately 2 times and 1.5 times faster as compared to TOP3D125MGCG and TOP3D125MGCGOMP, respectively. Similar observations can be found when run on systems 2 and 3 respectively (Fig. \ref{fig:cant_conv}).

Finally, to illustrate the scalability of the proposed GPU based framework, we conduct a case study by varying the number of degrees of freedom of the system (achieved by varying the discretization). We compare TOP3D125MGCG and TOP125MGCGCUDA. It is observed that both memory and time per iteration required for the proposed CUDA based topology optimization framework increases at a much slower rate as compared to the MGCG version. This indicates the superior scalability of the proposed approach.

\begin{table}[htbp!]
    \centering
    \caption{Comparison of performance of various codes in system 1}
    
    \label{tab:com_sys1}
    \begin{tabular}{|p{4.5cm}|p{3cm}|p{2cm}|p{3cm}|p{3cm}|}
        \hline 
            \textbf{Code} & \textbf{Discretization} & \textbf{Time per iteration} & \textbf{Memory requirement (GB) } & \textbf{No of iterations to converge} \\
        \hline
        
            TOP3D125 & 64x32x32 & 44.7s & 15.08 & 64\\
        \hline
        
            TOP3D125MGCG & 64x32x32 & 6.4s & 3.27 & 64\\
        \hline
        
            TOP3D125MGCGOMP & 64x32x32 & 2.6s & 2.60 & 64\\
        \hline
        
            TOP3D125MGCGCUDA & 64x32x32 & 1.0s & 0.04 (2.03) & 64\\
        \hline
        
            TOP3D125 & 128x64x64 & \multicolumn{3}{c|}{Out of memory}\\
        \hline
        
            TOP3D125MGCG & 128x64x64 & 44.9 & 3.98 & 49\\
        \hline
        
            TOP3D125MGCGOMP & 128x64x64 & 33.8s & 2.35 & 49\\
        \hline
        
            TOP3D125MGCGCUDA & 128x64x64 & 21.3s & 0.414 (2.79) & 49\\
        \hline
    \end{tabular}
\end{table}

    
        
        
        
        
            

\begin{figure}[htbp!]
        \centering
        \subfigure[]{
        \includegraphics[width = 0.46\textwidth]{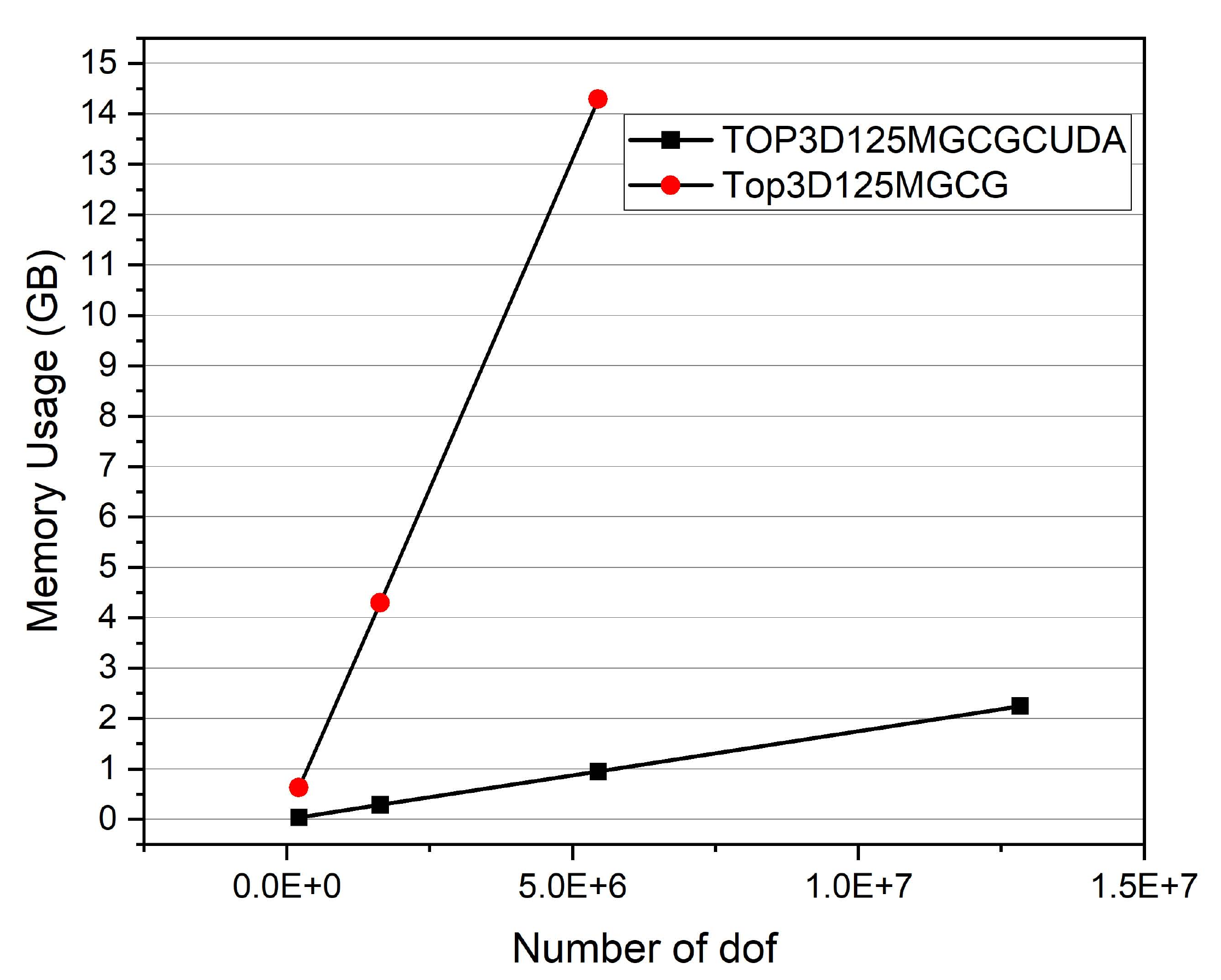}}
        \hspace{0.1mm}
        \subfigure[]{
        \includegraphics[width = 0.47\textwidth]{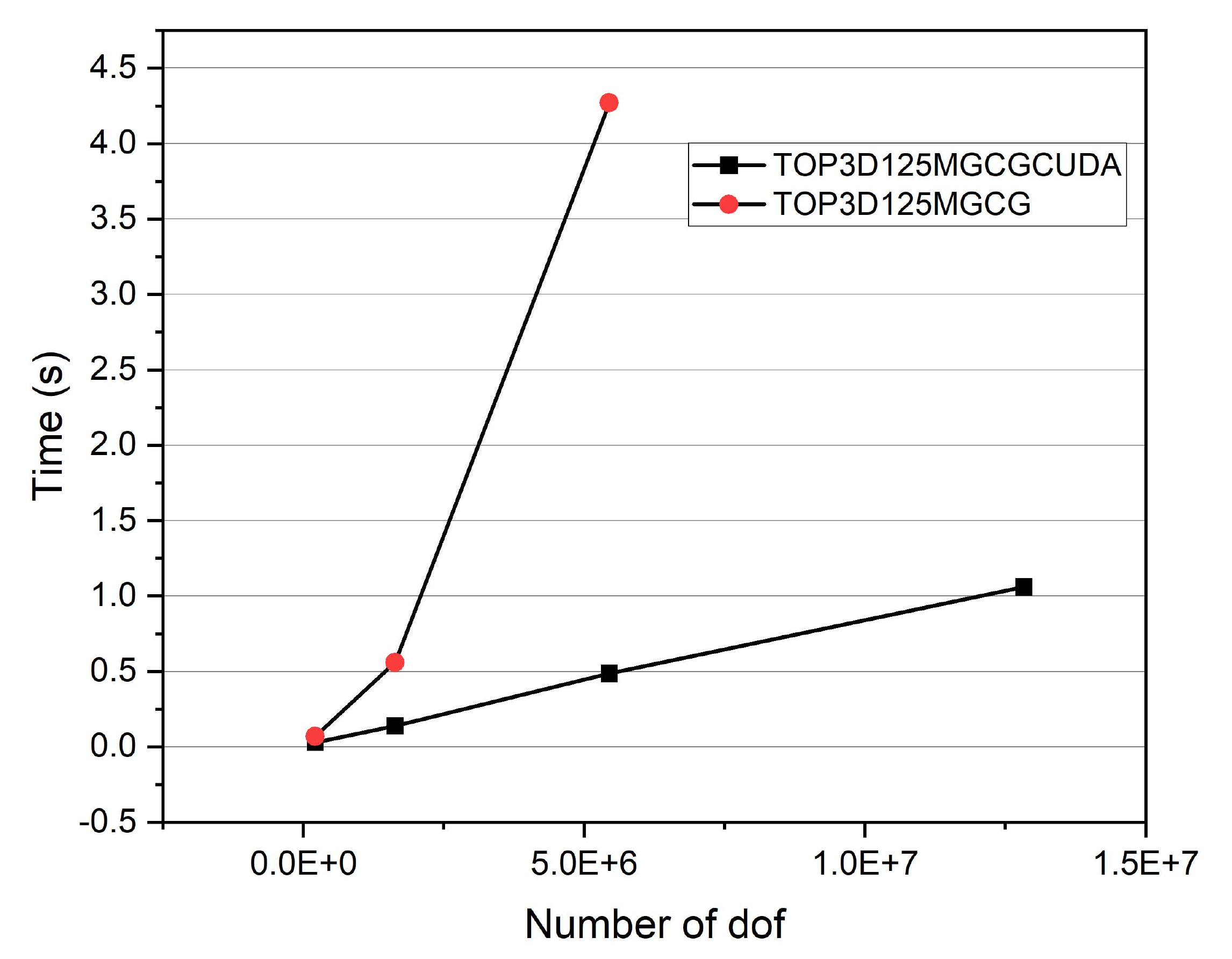}}
        \caption{Memory and time of computation comparison}
        \label{fig:time-memo-comp}
\end{figure}

\subsection{Example 2: Arch Bridge}

\begin{figure}[htbp!]
        \centering
        \subfigure[]{
        \includegraphics[width = 0.8\textwidth]{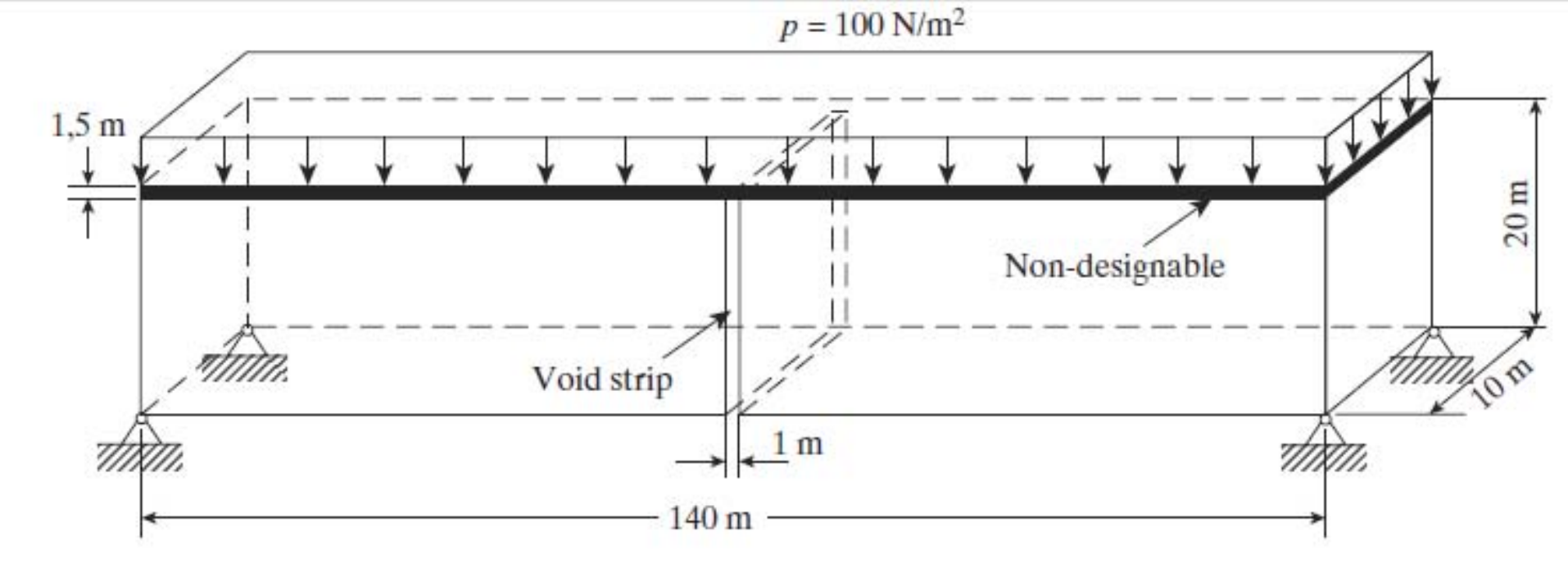}}
        \hspace{8mm}
        \subfigure[]{
        \includegraphics[width = 0.4\textwidth]{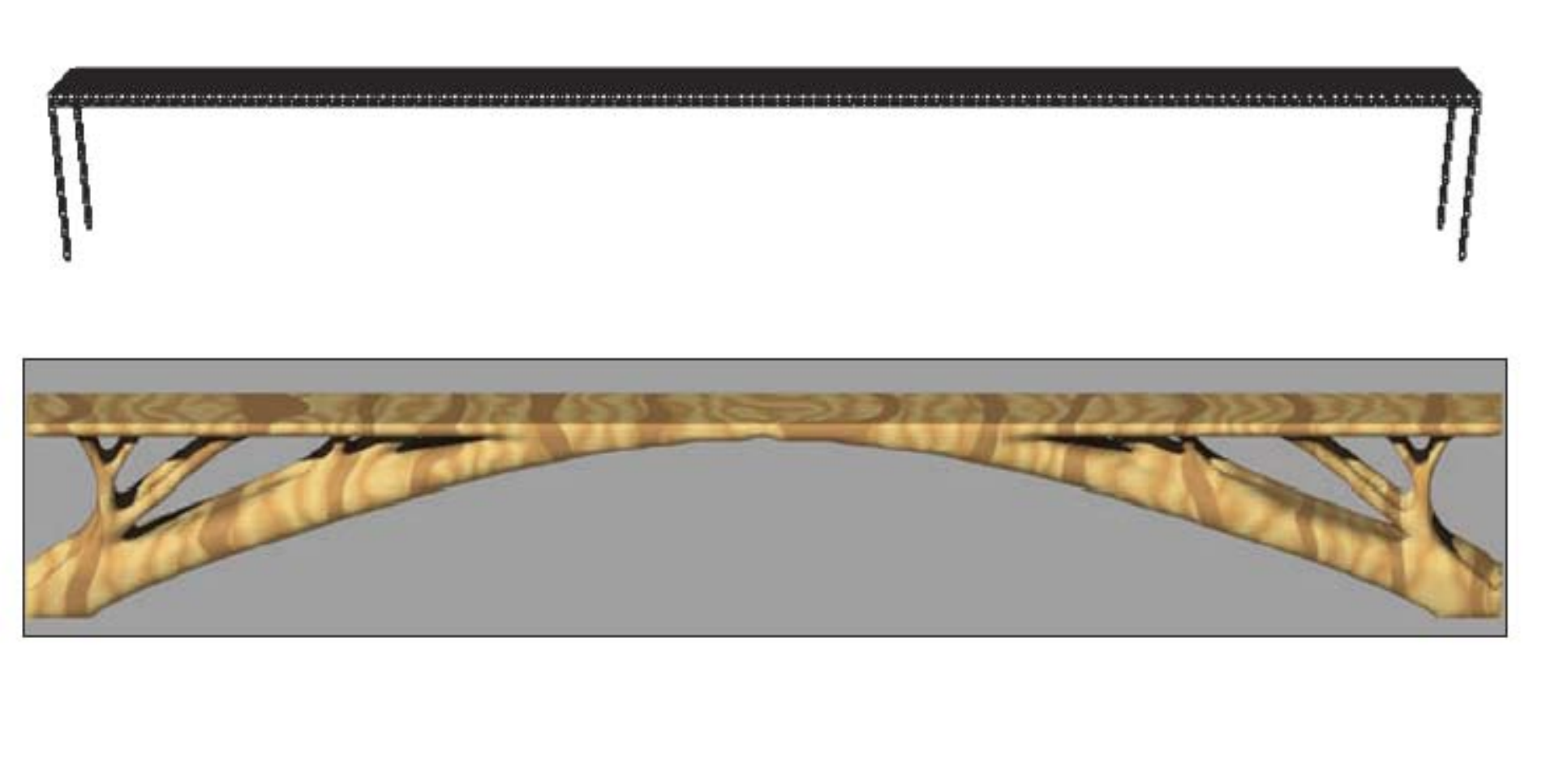}}
        \subfigure[]{
        \includegraphics[width = 0.4\textwidth]{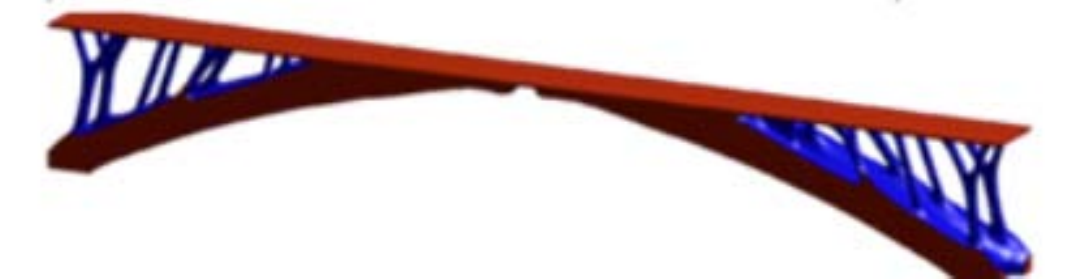}}
        \caption{Arch Bridge Example (a) Problem Definition (b) Initial Guess and solution from BESO (c) Solution from our code}
        \label{fig:Arch_1}
\end{figure}

As the second example, we illustrate the performance of the developed framework on passive void and passive solid region within design domain to obtain bridge like shapes similar to \cite{xie2014application}. Similar to previous example, we study two different cases. In the first case, we consider a $140m \times 10m \times 20m$ design domain. The top layer having $1.5m$ thickness is considered as passive solid region (non-design domain) as shown in Fig. \ref{fig:Arch_1}(a). Similarly, at the midway of $140m$ length a small void region of thickness $1m$ is treated as passive void. Boundary conditions and loading include simple support at four bottom corners and an uniformly distributed load (UDL) of $100 N/m^2$ on the top surface.

A discretization of $448 \times 32 \times 64$ is considered with a target volume fraction of 0.14. The resultant shape of compliance minimization problem is shown in Fig. \ref{fig:Arch_1}(c). The output is an arch bridge whereas an easy guess could be that of a bench kind of shape.
Compared to the output from \cite{xie2014application}, our result looks mostly similar; this validates the accuracy of the proposed approach for this problem. 

\begin{figure}[htbp!]
        \centering
        \subfigure[]{
        \includegraphics[width = 0.8\textwidth]{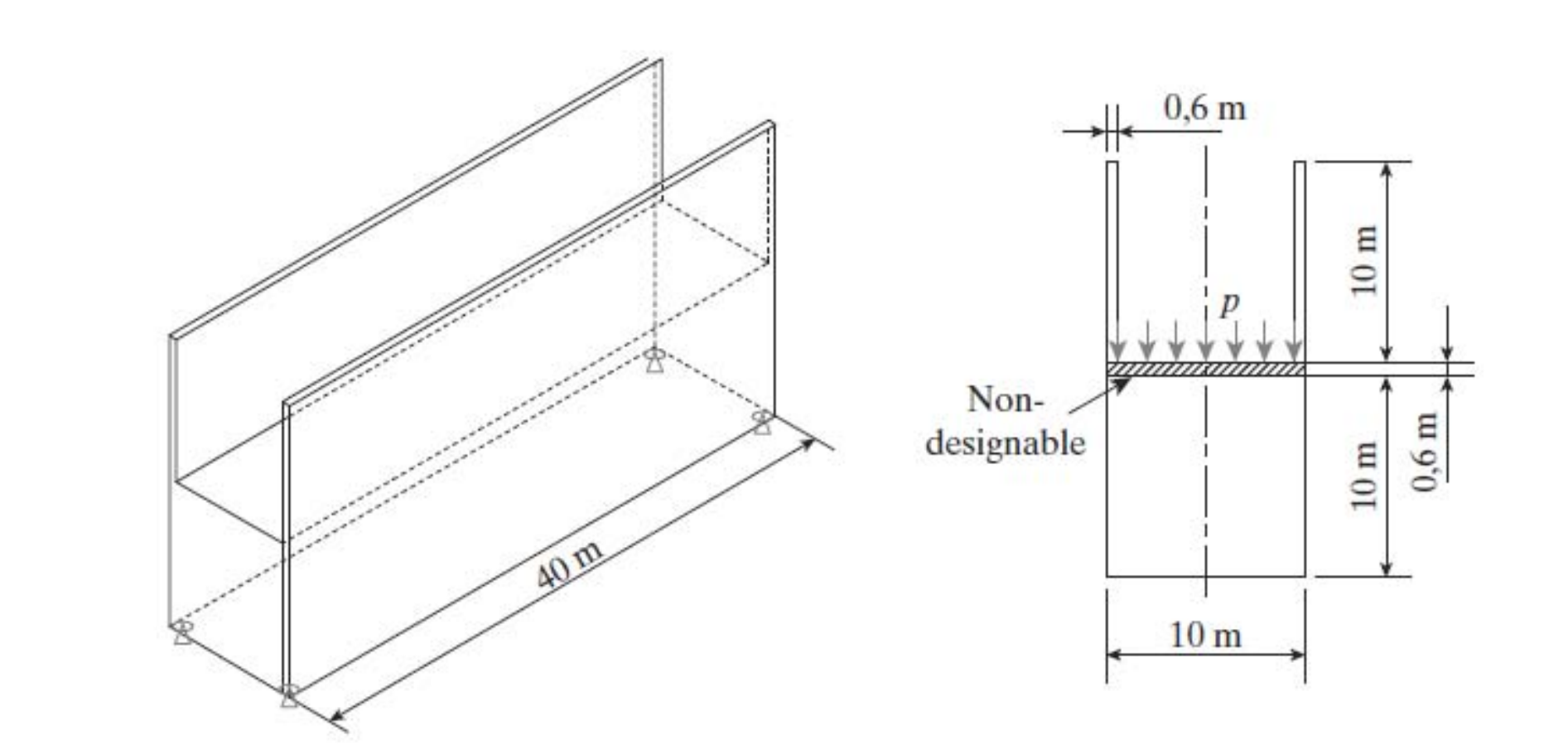}}
        \hspace{8mm}
        \subfigure[]{
        \includegraphics[width = 0.8\textwidth]{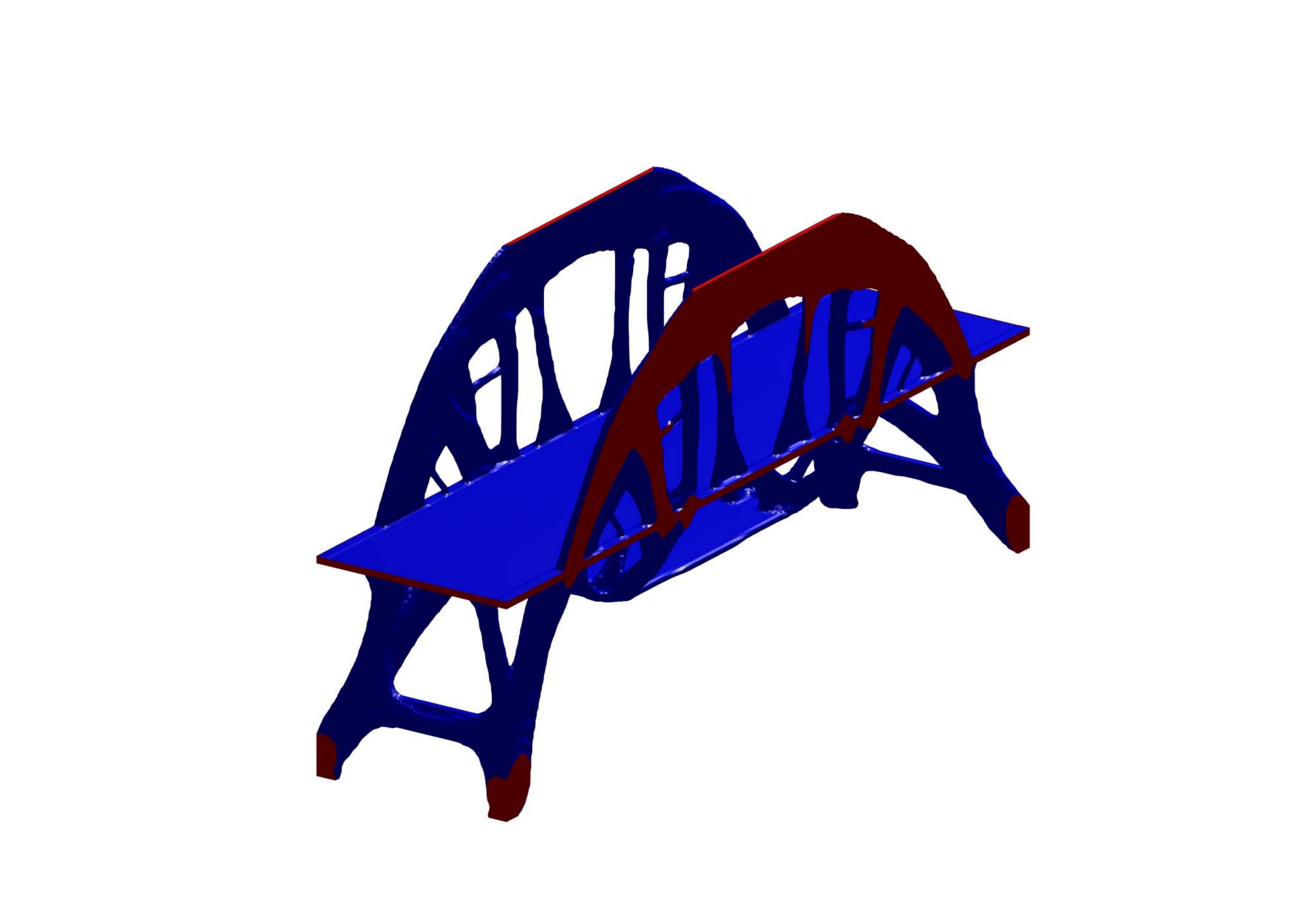}}
        \caption{Arch Bridge Example (configuration 2) (a) Problem Description (b) Minimum Compliance design}
        \label{fig:Arch_2}
\end{figure}

A second similar case is shown in Fig. \ref{fig:Arch_2} where the span of the design domain is $40m$. Carriageway width and height are respectively $10m$ and $20.6m$. At the mid-height, a non-designable (passive solid) layer of 0.6m thick is considered corresponding to the deck of the bridge. Also just above it, a void region of $40m \times  8.8m \times 10m$ is considered corresponding to space required for vehicular passage. The uniformly distributed loading on the deck and the boundary conditions are shown in Fig. \ref{fig:Arch_2}(a). The optimization is carried out with using a discretization of $256 \times 64 \times 128$ which accumulates to around 2 millions 8-noded 3D elements. The final shape after 49 iterations comes out to be an arch bridge as shown in Fig. \ref{fig:Arch_2}(c).

The summary of computing time and memory requirement of two type of arch bridges described above is summarized in Table \ref{tab:arch_bridge}. We observe that the proposed hybrid topology optimization framework is highly efficient both in terms of computational efficiency (26.27 and 224.18 s per iteration) and computational memory (0.486 GB and 1.14 GB) and can run even on a old system with 2 GB GPU memory. 

\begin{table}[htbp!]
    \centering
    \caption{Arch Bridge analysis details}
    
    \label{tab:arch_bridge}
    \begin{tabular}{|p{4cm}|p{3cm}|p{3cm}|p{3cm}|p{3cm}|}
        \hline 
            \textbf{Design domain dimension} & \textbf{Discretization} & \textbf{Total no. of iterations} & \textbf{Time per iteration} & \textbf{Avg. memory} \\
        \hline
            140m $\times$ 10m $\times $ 20m & 448 $\times$ 32 $\times $ 64  & 57 & 26.27 s & 0.486 GB\\
        \hline
            40m $\times$ 10m $\times $ 20m & 256 $\times$ 64 $\times $ 128  & 49 & 224.18 s & 1.14 GB\\
        \hline
    \end{tabular}
\end{table}

\subsection{Example 3: High Rise Building}
\begin{figure}[htbp!]
        \centering
        \subfigure[]{
        \includegraphics[width = 0.2\textwidth]{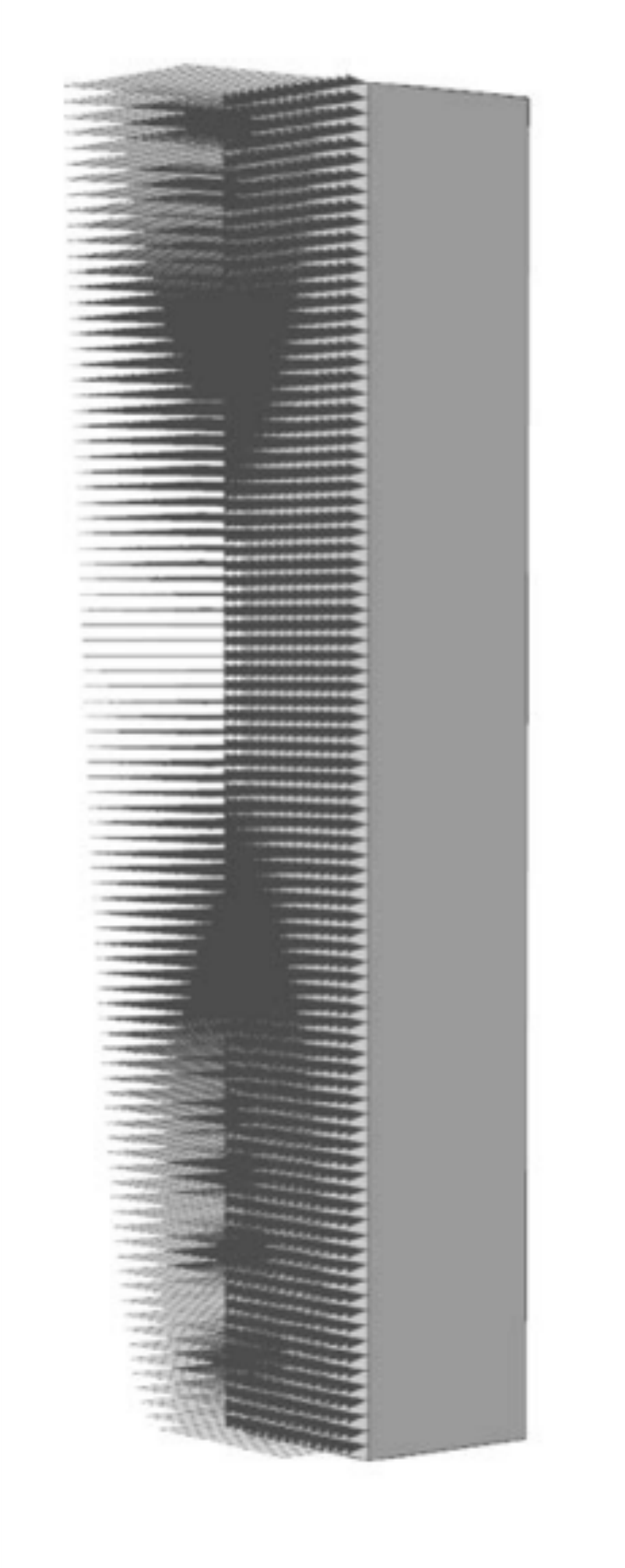}}
        \hspace{2mm}
        \subfigure[]{
        \includegraphics[width = 0.3\textwidth]{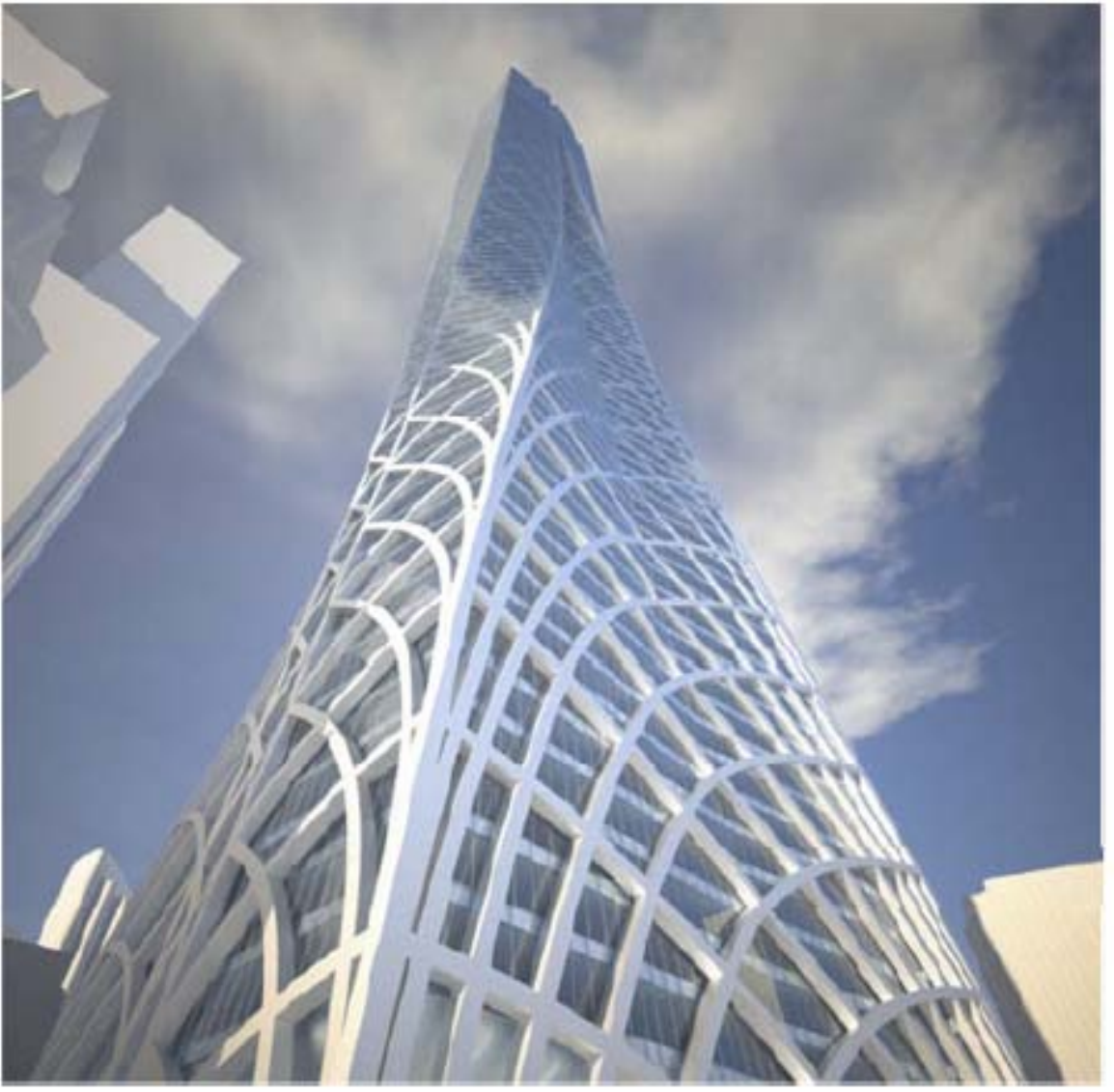}}
        \hspace{2mm} 
        \subfigure[]{
        \includegraphics[width = 0.3\textwidth]{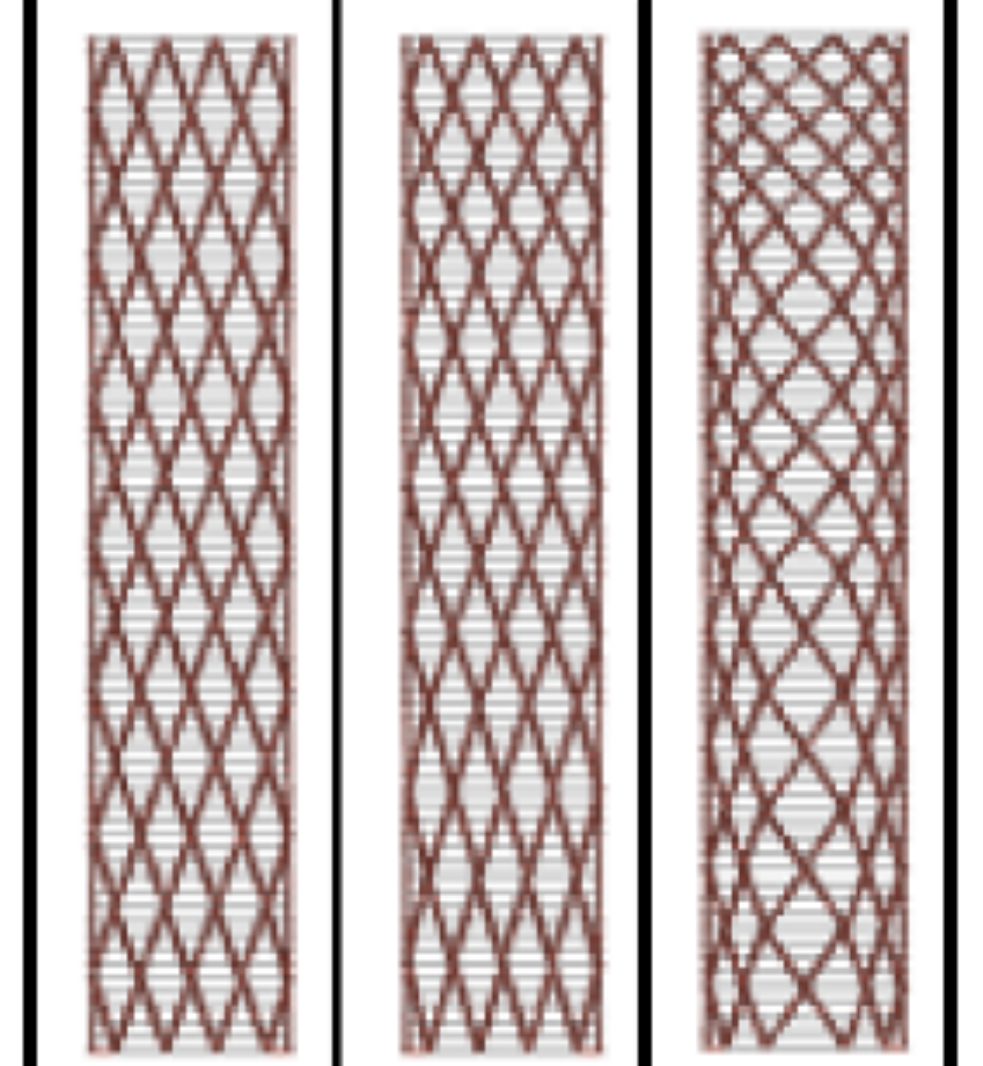}}
        \hspace{2mm}
        \subfigure[]{
        \includegraphics[width = 0.1\textwidth]{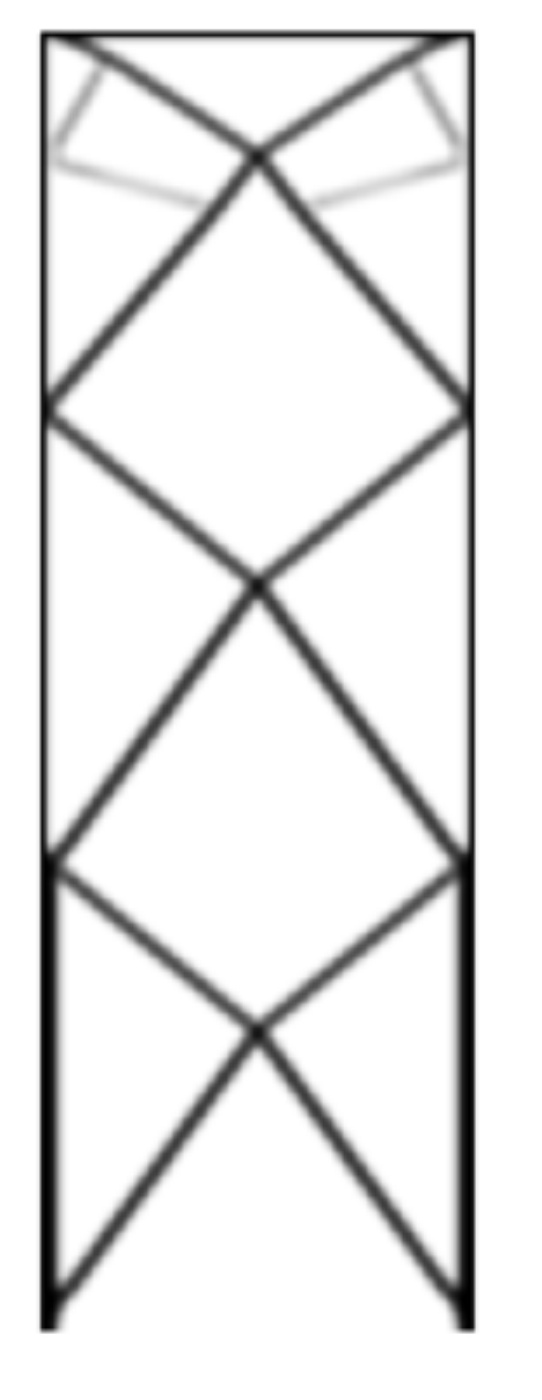}}
        \hspace{2mm}
        \subfigure[]{
        \includegraphics[width = 0.1\textwidth]{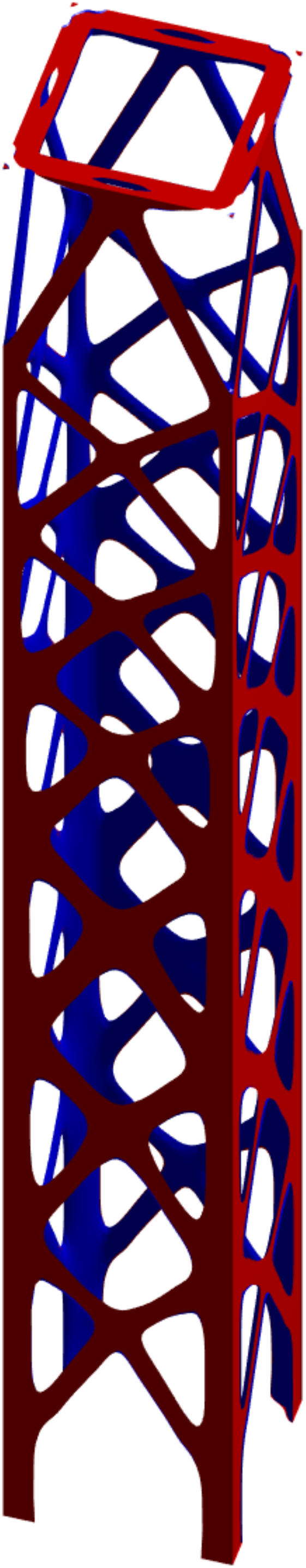}}
        \hspace{2mm}
        \caption{High Rise Building: (a) Problem Description \cite{cascone2021stress} (b) Concept Design of real world transit bay tower \cite{sarkisian2010organic} (c) Various Principal stress inspired Design from \cite{cascone2021stress} (d) Our 2D designs (e) 3D design}
        \label{fig:High_rise}
\end{figure}

In this example, a high rise tall building has been analyzed. The building is acted upon by the wind loading and the lateral load resisting frame system is to be obtained from minimization of compliance. The plan dimension of the building is $64m \times 64m$. The height to plan dimension ration $H/B = 4$. For simplicity, the base of the building is kept fixed and a constant magnitude of lateral loading is applied at all floor levels across the elevation of the building on one side of the building. A similar work is recently reported in \cite{cascone2021stress} in which the authors have obtained a diagrid pattern (Fig. \ref{fig:High_rise} (c)) of eccentric bracing system by aligning material along principal stress directions.

In our experiment, a discretization of $64 \times$ 64 $\times  256$ is considered. Subsequently, we have over 1 million design variables. The core of the building is kept hollow\footnote{for numerical stability, we consider a very small stiffness of $1/10^6$ }. A bracing system similar to \cite{cascone2021stress} has been obtained corresponding to a volume fraction of 0.12 in the active perimeter region. As for computational time, the proposed approach takes around 33s per iteration and 64 iteration to yield converged solutions. As for computational memory, the proposed approach require only 0.71 GB of GPU memory. The final optimized configuration is shown in Fig \ref{fig:High_rise}. 

\subsection{Example 4: Foot Bridge}
\begin{figure}[htbp!]
        \centering
        \subfigure[]{
        \includegraphics[width = 0.45\textwidth]{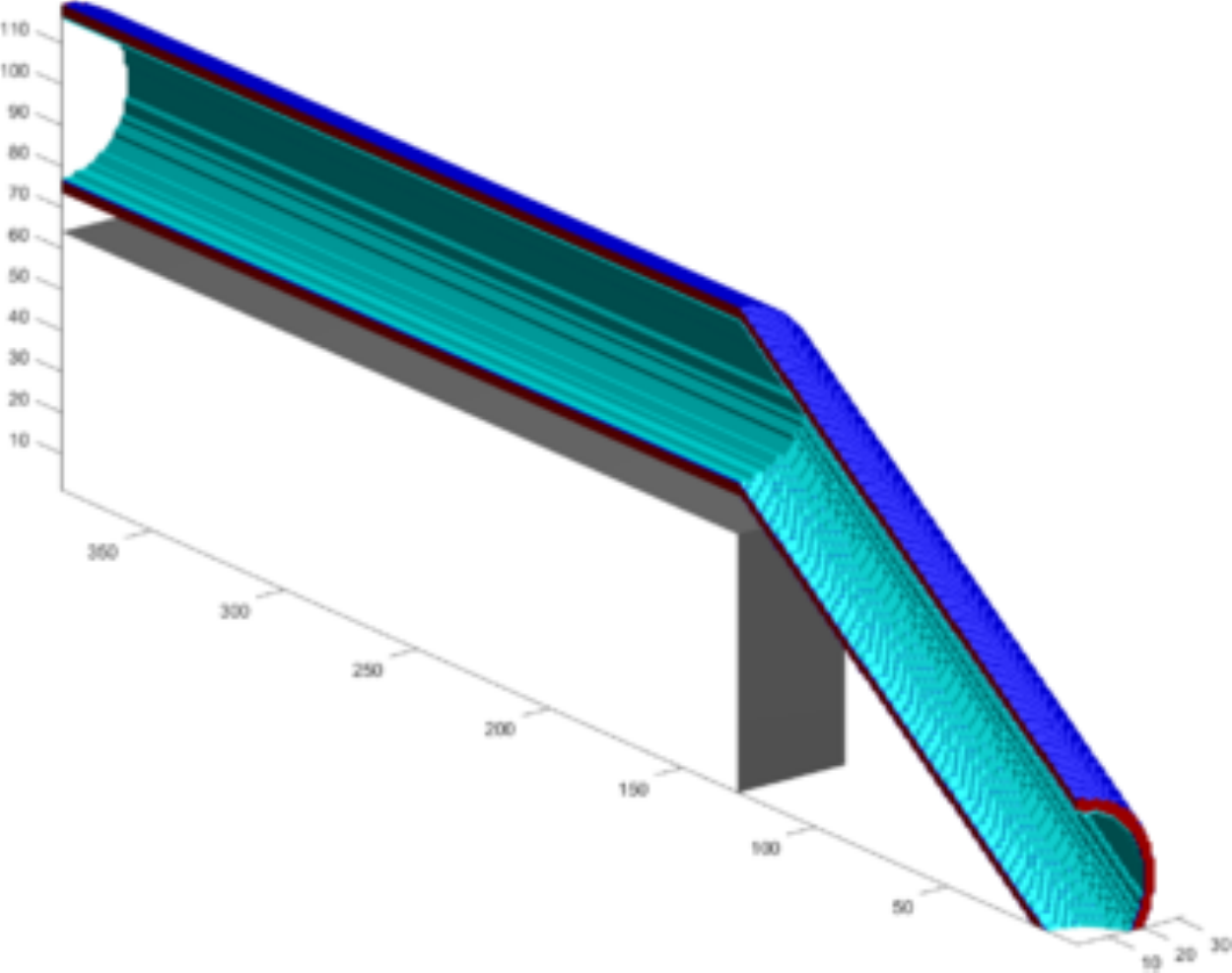}}
        \subfigure[]{
        \includegraphics[width = 0.5\textwidth]{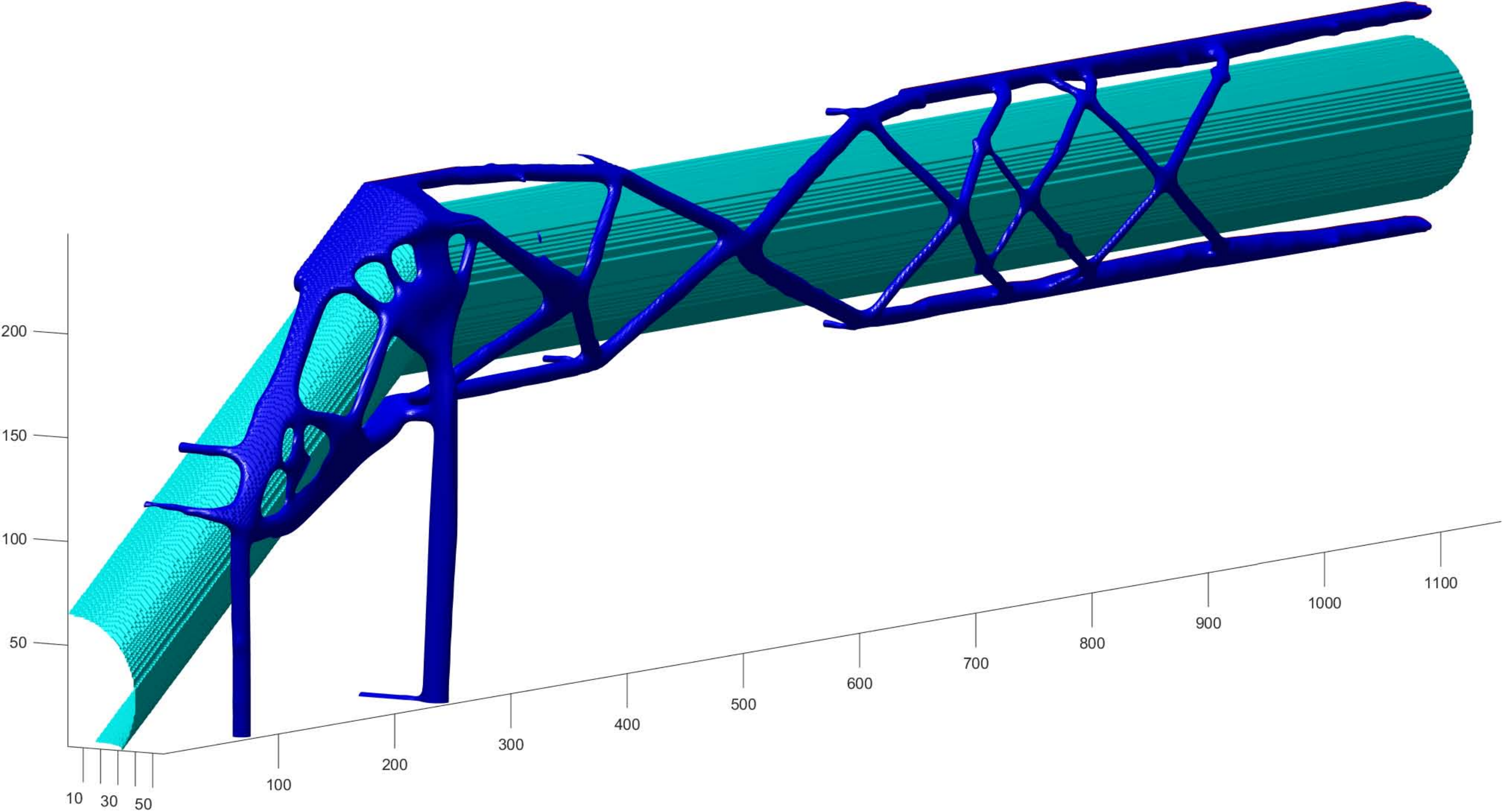}}
        \subfigure[]{
        \includegraphics[width = 0.9\textwidth]{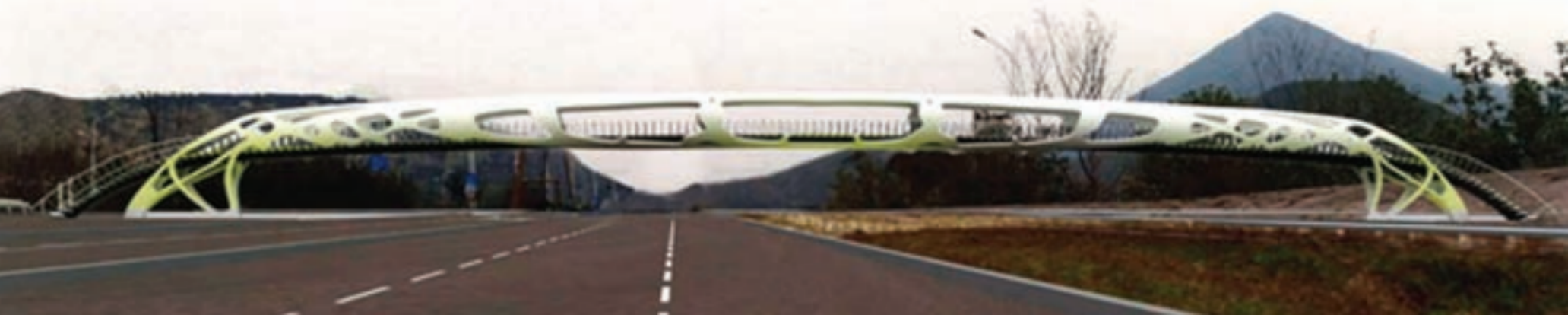}}
        \hspace{2mm}
        \caption{(a) Design Domain highlight(b) Optimized support system for the foot bridge obtained (c)Solution from literature \cite{xie2014application}}
        \label{fig:foot_bridge}
\end{figure}

As the fourth example, we optimize a foot bridge for gravity loading. Since gravity load is dependent on the amount of material present, the load varies with each iteration and complicates the optimization process. Subsequently, higher number of iterations is necessary to achieve converged solution. The design domain for the problem is shown in Fig. \ref{fig:foot_bridge}(a) wherein gray parallelepiped whose surface is the passive void region and internal light green tube is passive solid region. Dark blue layer represents the active region. Gravity loading proportional to volume of active and passive solid is applied in each optimization iteration and support structure for the tubular passage is obtained. A discretization of 1152 $\times $ 64 $\times$ 256 (which is around 58 millions of elements) is considered and optimization is carried out for a target volume fraction of 0.125 in the support structure region. The optimized configuration obtained using the proposed approach is shown in Fig. \ref{fig:foot_bridge}(b). We observe that the pattern is almost similar to an actual solution from literature \cite{xie2014application}. Overall the proposed approach converges in 49 iterations and requires 3.924 GB GPU memory as detailed in Table \ref{tab:foot_bridge}.

\begin{table}[htbp!]
    \centering
    \caption{Foot Bridge analysis details}
    
    \label{tab:foot_bridge}
    \begin{tabular}{|p{4cm}|p{3cm}|p{3cm}|p{3cm}|p{3cm}|}
        \hline 
            \textbf{Design domain dimension} & \textbf{Discretization} & \textbf{Total no. of iterations} & \textbf{Time per iteration} & \textbf{Avg. memory} \\
        \hline
            180m $\times$ 10m $\times $ 40m & 1152 $\times $ 64 $\times$ 256  & 49 & 2850 s & 3.924 GB\\
        \hline
    \end{tabular}
\end{table}

\section{Homogenization enhanced hybrid topology optimization approach} 
In the previous section, we illustrated the performance of the proposed approach in solving four topology optimization problems. While we illustrated the efficacy of the proposed approach, it was observed that the proposed framework has memory cost in Gigabytes (GB) to the GPU which may restrict the applicability of the method to realistic scenarios. One way to address this issue is to introduce a homogenization step in the interpolation phase (from coarser to finer grid) of the V-cycle \cite{alcouffe1981multi,hoekema1998multigrid}. The basic idea here is to represent the density of an element by computing the average value of eight finer elements; this reduces the storage requirement of stiffness matrices of each element in the coarser grids in the V-cycle part of the algorithm.  Thus, this method is seen to perform significant better in memory cost to GPU and has improved convergence. The comparison of memory requirements in a benchmark cantilever beam is summarized in Table \ref{tab:homogenization}. We observe that the use of homogenization reduces the memory requirement by 40-50$ \% $ depending on the problem size and other factors. 

\begin{table}[!hbt]
    \centering
    \caption{Comparison of performance of homogenized multigrid approach}
    
    \label{tab:homogenization}
    \begin{tabular}{|p{4cm}|p{3cm}|p{2cm}|p{3cm}|p{3cm}|}
        \hline 
            \textbf{Algorithm} & \textbf{Discretization} & \textbf{Time per iteration} & \textbf{Memory (avg. in iteration)} & \textbf{No of iterations to converge} \\
        \hline
            2015 and 2017 GPU TO approach \cite{wu2015system} & 64x32x32 & 1s & 40 MB & 64\\
        \hline
            Our Homogenized MG Code & 64x32x32 & 3s & 19 MB & 64\\
        \hline
            2015 and 2017 GPU TO approach \cite{wu2015system} & 128x64x64 & 21.6s & 414 MB & 49\\
        \hline
            Our Homogenized MG Code & 128x64x64 & 22.4s & 243 MB & 49\\
        \hline
    \end{tabular}
\end{table}

\begin{figure}[htbp!]
        \centering{
        \includegraphics[width = 0.15\textwidth]{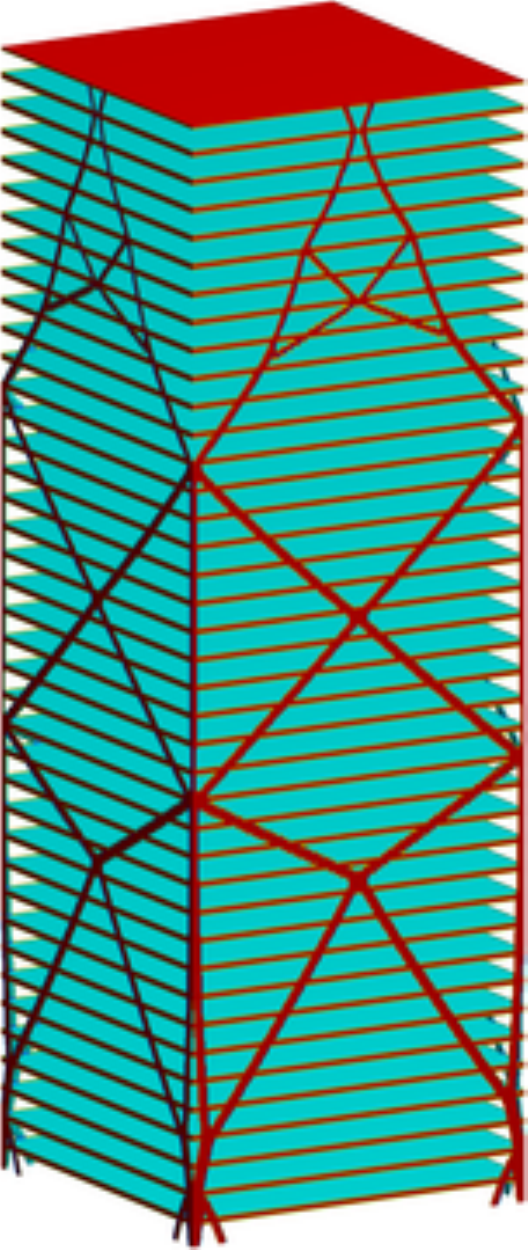}}
        \caption{Optimized configuration for the high-rise building problem}
        \label{fig:Tall_B_Homo}
\end{figure}

Finally, we consider a high-rise building problem having plan area of $54m \times 54m$. The height of the building is $162m$. The objective here is to illustrate the capability of the proposed algorithm in solving a highly heterogeneous system. We have considered the system to be subjected to parabolic loading profile \cite{cascone2021stress}. We consider the floors to be of concrete with Elastic modulus of 25000 MPa. The objective here is to minimize the structural compliance; however, unlike the previous examples, we keep a track on on the maximum top story drift following the Eurocode for building \cite{standard2010eurocode,standard2004eurocode}. The analysis of this structure is carried to calculate displacement subjected to the lateral loading. Table \ref{tab:homogenization_comp} summarizes the performance of standard Galerkin scheme and homogenization based scheme on this system. We observe that the homogenization approach has around 43\% less cost to GPU memory and its iterations takes less time due to faster convergence of the MGCG iterations per each optimization redesign loop. The optimized topology obtained is shown in Fig. \ref{fig:Tall_B_Homo}. We observe that the top floor drift for the optimized configuration is 0.026m which is significantly lower than the allowed threshold.

\begin{table}[htbp!]
    \centering
    \caption{Homogenization approach on high rise building}
    
    \label{tab:homogenization_comp}
    \begin{tabular}{|p{4cm}|p{3cm}|p{3cm}|p{3cm}|p{3cm}|}
        \hline 
            \textbf{Approach} & \textbf{Discretization} & \textbf{Total no. of iterations} & \textbf{Time per iteration} & \textbf{Avg. memory} \\
        \hline
            Standard (Galerkin) & 64 $\times $ 64 $\times$ 320  & 84 & 102.58 s & 1.08 GB\\
        \hline
            Homogenization & 64 $\times $ 64 $\times$ 320  & 81 & 86.37 s & 0.62 GB\\
            
        \hline
    \end{tabular}
\end{table}

\section{Conclusion}\label{sec:Conl}
In this work, we have proposed a new GPU enhanced hybrid topology optimization framework for structural compliance minimization. The proposed approach replaces the primary bottleneck associated with direct solution of state space system with an efficient mutltigrid conjugate gradient method. The proposed approach utilizes both CPU and GPU; in specific, all tedious arithmetic computations have been shifted to the highly multithreaded cores of modern GPU based on CUDA architecture. Additionally, we have utilized a simple homogenization scheme within the proposed approach that drastically reduces the memory requirement and improves the computational efficiency. Overall, the proposed approach is highly efficient and easily scales to systems having millions of degree of freedoms.

Two version of our framework has been developed and implemented - one for CUDA GPU based system (TOP3D125CUDA) and another for purely CPU system (TOP3D125OMP). Several examples are solved to illustrate the performance of the proposed approach. In a standard cantilever beam benchmark problem, our CUDA based algorithm is about two times faster and consumes 7-8 times lower memory compared to contemporary efficient implementation. Similarly the OMP version is also 1.5 times faster, although it consumes similar memory compared to the state-of-the-art implementations. It is worthwhile to note that the increased efficiency is achieved without any compromise in the accuracy. One of the primary feature of the proposed approach is its scalability. It easily scales to millions of degrees of freedom. The proposed framework consumes only 1.1 GB of GPU memory and computational time of around 4 minutes for solving the arch-bridge problem having over 2 million degrees of freedom. Additionally a further extension of our algorithm to incorporate a homogenization scheme is shown to reduce the memory requirement by around 40-50\% in the cantilever benchmark problem. Also with homogenization approach in place a rough calculation shows that a system with 100 millions of 3D elements can fit into the memory of a standard 8GB GPU for analysis. Faster convergence of the homogenization approach in the high rise building having high heterogeneity has also been demonstrated. Overall our framework produced excellent results across various examples with significant efficiency in computation and memory requirement.

Despite the excellent performance, the proposed framework has certain limitations. Firstly, in its current form, the proposed framework is applicable to structured mesh. For unstructured mesh, an additional step involving dividing the problem domain into sub-regions having similar discretization is necessary. Secondly, the developed CUDA based framework will only work with NVIDIA GPUs. This is not a significant limitation when one considers the fact that NVIDIA currently has over 77\% \cite {otterness2020amd} market share. Also, the developed OMP version of the framework is universal and will work on any windows based system. In future, some of these limitations will be addressed.

\section*{Acknowledgements}
We acknowledge the financial support received from IIT Rookee in form of MHRD, Govt. Of India, fellowship.
\section*{Conflict of interests} The authors declare that they have no conflict of
interest.
\section*{Replication of results} Matlab codes as well as the dependent CUDA C routines for the examples shown in this work will be available freely in GitHub once the paper is accepted.

\appendix

\section{Three different System Configurations Used for Testing}

\begin{table}[htbp!]
    \centering
    \caption{Specification of systems used for testing}
    
    \label{tab:sys-specs}
    \begin{tabular}{|p{2.5cm}|p{4.5cm}|p{4.5cm}|p{4.5cm}|}
        \hline 
           \textbf{Component} & \textbf{System - 1} & \textbf{System - 2} & \textbf{System - 3} \\
        \hline
            \textbf{CPU} & 2 x Intel Xeon Gold 5218 & 2 x Intel Xeon X5660  & Intel i7 9750H\\
        \hline
            \textbf{GPU} & NVIDIA QUADRO RTX4000 with 8GB GDDR6X VRAM & NVIDIA QUADRO K620 with 2GB GDDR3 VRAM  & NVIDIA GeForce GTX 1650 with 4GB GDDR5 VRAM\\
        \hline
            \textbf{RAM} & 192 GB DDR4 & 48 GB DDR3  & 16 GB DDR4 \\
        \hline
            \textbf{OS} & Windows 10 x64 20H2 & Windows 10 x64 20H2  & Windows 10 x64 20H2\\
        \hline
            \textbf{MATLAB} & R2020b & R2020b  & R2020b \\
        \hline
    \end{tabular}
\end{table}


\end{document}